# Flexibility and Invariance of Object Representations in the Human Brain

**Julien Dirani[1,2,3], Shankar Chawla[2], Leila Wehbe[*1,3], Bradford Z. Mahon[*1,2,4]**

1 Neuroscience Institute, Carnegie Mellon University, USA
2 Department of Psychology, Carnegie Mellon University, USA
3 Department of Machine Learning, Carnegie Mellon University, USA
4 Department of Neurosurgery, University of Rochester Medical Center, USA
* Co-senior authors

**Abstract**

The human brain represents objects in a way that is both invariant across instances and flexible enough to support different contexts and tasks. Yet how the brain reconciles these demands, holding an object's representation stable while flexibly reshaping it to meet the current context, remains unknown. Using fMRI during naturalistic movie viewing we investigated how the same objects are represented when they are passive scene elements versus targets of goal-directed actions. Action targets engaged a parietal action network centered in the supramarginal and postcentral gyri, while passive objects recruited a distributed occipito-temporal network involved in visual object recognition. Within context-selective networks, representational geometry showed a double dissociation: target objects were organized by action affordance and hand posture affordance dimensions, while passive objects aligned with semantic dimensions. The visual structure of object representations, by contrast, was invariant across contexts. Searchlight analyses further showed that this dissociation reflected the relative engagement of networks of regions rather than the strict presence or absence of each representational format. Flexibility and invariance are not opposing principles of neural representation, but operate simultaneously at different levels of a common, distributed representational system.

### Highlights

- Object representations remap according to contextual role
- Action targets recruit a parietal network organized by affordance structure
- Passive objects recruit an occipito-temporal network organized by semantic structure
- Visual structure anchors a stable, context-invariant object representation
- Representational flexibility and invariance coexist at separable levels

## INTRODUCTION

A fundamental ability of the human brain is to represent objects in a way that is both invariant across instances and input modalities, while flexible enough to support ever-changing task demands and environmental layouts[1,2]. Object representations are high dimensional, both functionally in that they involve separable types of information, and anatomically, in that they are distributed over dissociable brain networks[3–7]. While there have been major advances in understanding how object geometries are represented in the brain[5–7], it remains unknown how the brain maintains stable representations of objects while flexibly reshaping them to support different contexts and tasks.

Goal-directed action constantly reshapes how objects are processed and which of their many representational dimensions are relevant, both when we ourselves act on them, and when we observe others doing so. Consider observing someone preparing a meal: as they reach for a glass of water, its volumetric and spatial dimensions are highlighted in support of the ongoing action. As they set it down and pick up a knife, the glass becomes a passive element of the scene, no longer relevant to interpreting the ongoing action. The knife, now the target, may be represented along the dimensions most relevant to its current role, such as the grip it affords, and the actions it supports. As the scene unfolds and objects shift between contextual roles, some representational dimensions are prioritized while others are deemphasized. For these reasons, object-directed action provides a powerful lens with which to study how complex neural systems support context-based object representations. We hypothesize that some dimensions of an object's representation remain invariant to changes in context while others are flexibly modulated by context.

The dimensions along which object representations are flexible, and those that remain invariant, are shaped by the fundamental constraints on the system's architecture. At the highest level, a key constraint is a division of visual processing between the ventral occipito-temporal "what" pathway[4,8,9] and the dorsal occipital-parietal "how/where" pathway, which originates in subcortical and early visual structures[10–12]. The ventral pathway supports the computation of a stable perceptual representation of objects, in support of visual object recognition, visual categorization, and access to semantic knowledge[4,13]. Lateral occipital cortex plays an important role in object recognition and shape processing[8], while ventral temporal cortex more broadly supports object knowledge[4], including texture[14], color[15], and shape[9] as well as inferred physical attributes such as object weight[16]. The dorsal visual pathway projects cortically through dorsal occipital cortex and subcortically into the intraparietal sulcus (IPS), and supports real time analysis of objects in support of volumetrically and biomechanically calibrated grasping[17]. The dorsal stream supports the spatial and visuomotor transformations necessary for real-time planning, executing and updating of object-directed actions[13,17], including actions of the hand, arm, and eye, and egocentric representations of space[17] and retrieval of spatial context[18].

The dorsal visual pathway is one network within a broader set of separable parietal systems, including a parieto-premotor pathway supporting visually guided reaching and grasping[19,20], a parieto-prefrontal pathway for spatial working memory[21], and a parieto-medial-temporal pathway for spatial navigation[22]. *Why* an object is being grasped and *how* an object will be manipulated after it is grasped are not computed by dorsal stream systems. The 'why' behind actions depends on the integration of a semantic interpretation of the visual scene with an action goal, a process that is mediated by the ventral stream in interaction with frontal regions[2,22,23]. The inferior parietal cortex, and specifically the supramarginal gyrus, is a key region that supports both action production and observation[24–29], but which is not part of the dorsal visual pathway. Indeed, the supramarginal gyrus of the dominant inferior parietal lobule has rich interconnectivity to posterior middle temporal regions long implicated in the representation of object concepts and object function[30], as well as ventral occipito-temporal regions that support grasp-relevant representations of objects, such as texture, material properties and weight[30,31].

There is substantial evidence across a range of methods indicating that task and context modulate the representation of objects at both cognitive and neural levels[3,32–37]. Behaviorally, the object properties that are emphasized in perception are not fixed properties of the stimulus but are modulated by task- and context-sensitive top-down signals[38–40]. At the neural level, there is substantial evidence that top-down influences, including attention and task demands, modulate the strength of object-evoked responses[41–47]. fMRI studies using naturalistic movie viewing have shown that sustained attention to a behaviorally relevant category modulates the representation of voxels across occipito-temporal and fronto-parietal cortex toward attended targets[48]. In addition, explicitly accounting for task semantics significantly improves the prediction of object categories from MEG data[49]. Yet how, and through which neural systems, the brain balances flexible and invariant object representations as an object's role shifts across moment-to-moment changes remains unexplored. Testing this requires holding the objects themselves constant while varying their contextual role, which prior work has not done.

Here we use a naturalistic fMRI paradigm in which participants viewed videos of everyday activities in the kitchen, filmed from a first-person perspective. The videos were selected such that a common set of objects appears in dissociable contextual roles: as passive elements of the visual scene, or as targets of goal-directed actions. We leverage the rich, ecologically valid structure of naturalistic stimuli to vary the contextual relevance of objects to ongoing actions while holding object exemplars and the experimental task constant. Using voxel-wise encoding models and representational similarity analysis (RSA), we characterize not only which cortical regions are differentially recruited depending on an object's contextual role, but also the extent to which separable dimensions of object representations are modulated across contexts. To characterize cortical recruitment, we first fit voxel-wise encoding models using five feature spaces: passive objects, target objects, action labels, hand synergy weights, and motion energy. While passive and target objects are the primary feature spaces of interest, action labels and hand synergies were included to partial out the contributions of co-occurring actions and hand configurations. Motion energy was also included as a low-level visual covariate.

Within the networks most strongly encoding objects in each context, we test which representational dimensions dominate object representations. We test four hypothesis-driven Representational Dissimilarity Matrices (RDMs) against neural representational patterns: visual similarity, semantic similarity, action affordances, and hand posture affordances. Visual and semantic dimensions capture object appearance and conceptual structure, respectively. Action affordances reflect the distribution of actions an object supports, and hand posture affordances reflect the characteristic hand configurations recruited during its manipulation. Because object coefficients were estimated jointly with frame-by-frame action and hand configuration features, any alignment with affordance structure reflects object-level information while minimizing the influence of co-occurring motor variables.

## RESULTS

### *Overview*

Healthy participants watched videos of everyday cooking activities, with all videos acquired via a camera that was mounted on the agent's head. Videos were selected to represent a diversity of both actions and objects, with objects appearing across multiple actions and contextual roles. For instance, a cup may be washed in the sink, dried with a towel, or be present as a passive element on the counter. Frame-by-frame annotations were produced by trained human annotators and model-based feature extraction and used to construct the five feature spaces described above: passive objects, target objects, actions, hand synergy weights, and motion energy (***Figure 1***). To characterize the cortical regions supporting each feature space, we examined their distribution using product-measure split scores[50]. Voxel-wise contrasts then identified brain networks preferentially encoding objects in each context. Representational geometries were examined both within networks maximally tuned to each context and more broadly via searchlight RSA, allowing us to ask both where and how object representations are modulated by action context.

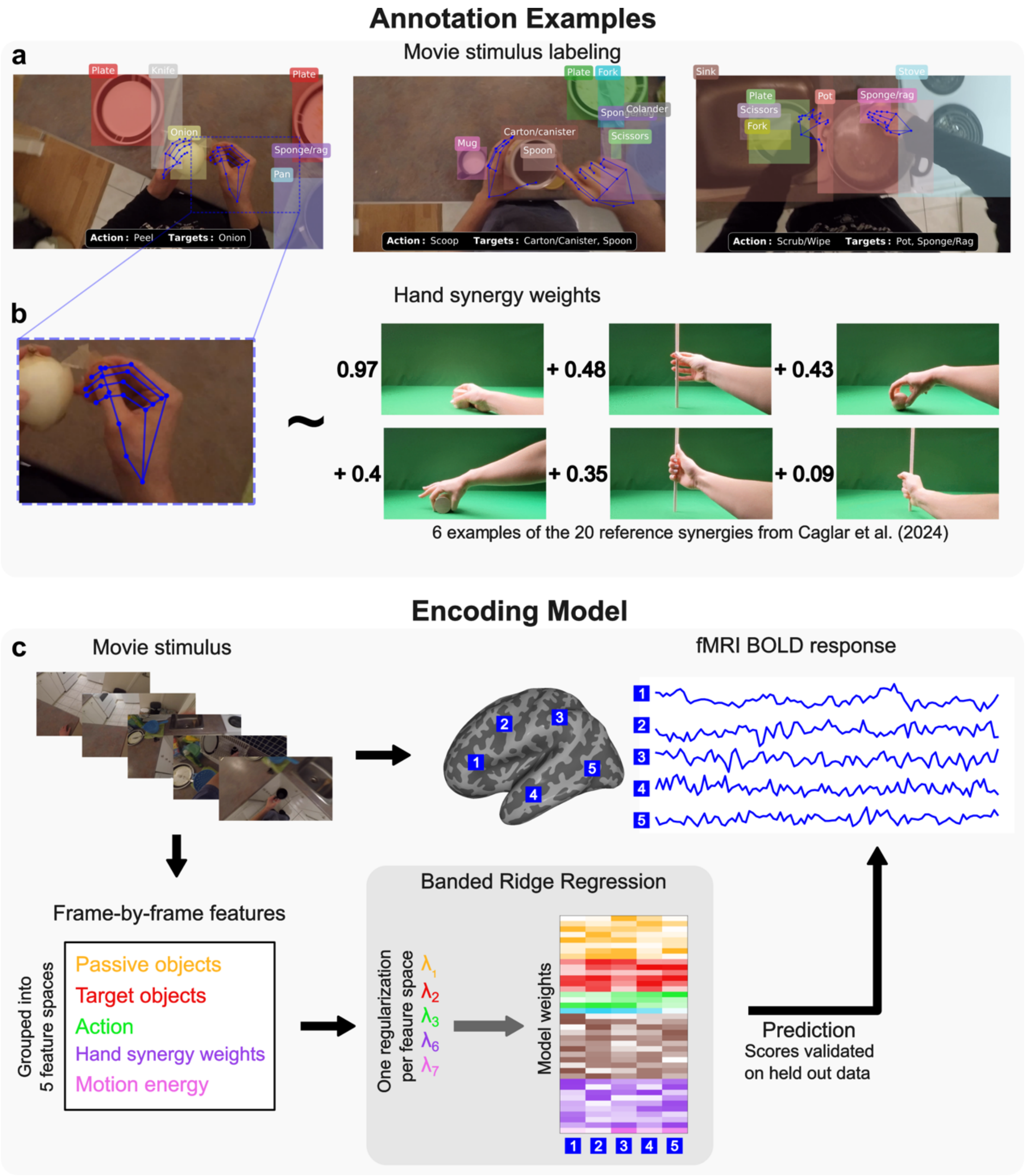

**Figure 1. Annotation of the first-person movie stimulus and overview of the encoding model.** First-person cooking video stimuli were annotated frame-by-frame and used to construct features that were fit to fMRI BOLD responses via banded ridge regression. **(a)** Three representative frames from the movie stimulus illustrate the annotation protocol. Frames were labeled for all visible objects (objects included tools, knobs/handles, food items, appliances), the ongoing action, and the target object(s) of the action. **(b)** The zoomed inset shows hand skeletal landmarks extracted from a single frame. Each observed hand configuration was decomposed into a weighted linear combination of 20 reference kinematic synergies from Caglar et al. (2025). Six example synergies are shown alongside their weights for the depicted hand posture, illustrating how hand skeletal landmarks are represented as a rotation- and translation-invariant description of hand configuration. **(c)** Frame-by-frame features were organized into five feature spaces (passive objects, target objects, action labels, hand synergy weights, and motion energy) and used as inputs to a voxel-wise banded ridge regression encoding model. Banded ridge regression assigns an independent regularization parameter to each feature space, allowing their relative contributions to voxel-wise BOLD predictions to be estimated jointly. Models are fitted within subjects where weights were learned on training runs and performance was evaluated as $R^2$ on a held-out test run.

### *Passive and target objects engage dissociable ventral and dorsal cortical networks*

The voxel-wise encoding model explained significant variance across a network spanning occipital, temporal, and parietal regions, with limited frontal involvement (**Figure *2***). Group-level effects were robust, with an observed mean t-value of 3.63 (SD = 1.20, range = 2.45–19.48), corresponding to Cohen's d = 1.48 and 93% power. Subject-level significance maps, derived independently via permutation testing, showed strong spatial overlap with the group-level results (Supplemental **Figure S2**). The banded ridge encoding model approach[50] allowed us to decompose the contribution of the five feature spaces into voxel-wise predictions of BOLD response (**Figure 3**).

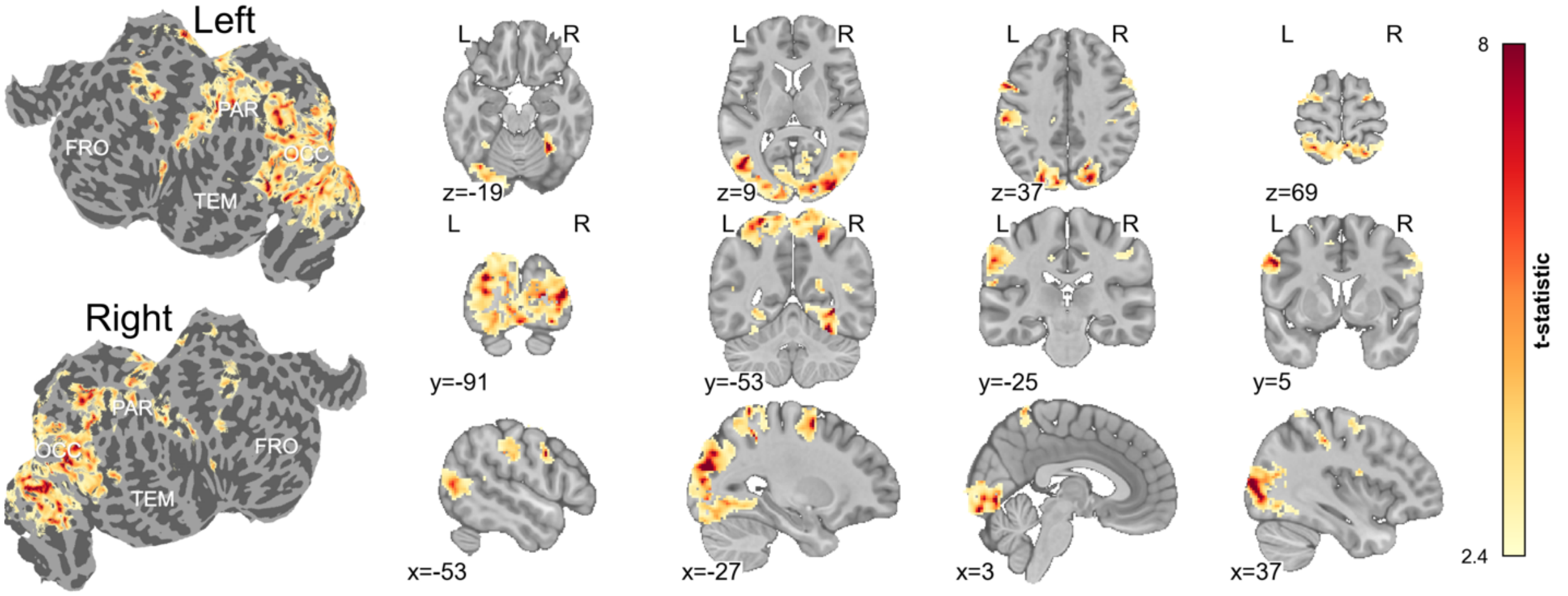

**Figure 2. Group-level coverage of the voxel-wise encoding model.** Group-level t-statistic map for voxel-wise model performance, testing for an $R^2 > 0$, shown on inflated cortical surfaces (left and right hemispheres; OCC: occipital, PAR: parietal, TEM: temporal, FRO: frontal) and on axial, coronal, and sagittal slices of the MNI template. Voxels not surviving FDR correction ($q < 0.05$, minimum cluster extent 10 voxels) are kept transparent. Model performance was significant bilaterally across occipital, temporal, and parietal regions, with some limited frontal involvement.

Action labels showed the strongest overall contributions of any feature space, with highest explained variance in bilateral middle occipital gyrus, bilateral cuneus, bilateral superior parietal lobule, left superior occipital gyrus, and bilateral lingual gyri. The strong prevalence of this effect is consistent with the temporally extended nature of actions, which aligns well with the slow timescale of the BOLD response.

Passive objects explained variance most strongly in bilateral middle and inferior occipital gyri and bilateral lingual gyri, with additional contributions in the cuneus and left fusiform gyrus, consistent with a ventral visual network supporting object identity and recognition[4,8,9]. Target objects, while explaining variance in fewer voxels, showed a notably distinct spatial profile: alongside occipital contributions shared with passive objects, target objects recruited somatosensory regions in the left postcentral gyrus, as well as the left supramarginal gyrus of the inferior parietal lobule. These regions have been consistently implicated in somatosensory processing, complex object-directed manipulation, and integration of semantic representations with visuomotor processing required of object-directed action[51,52]. Prior work has argued the supramarginal gyrus supports the recombination of an elementary vocabulary of motor synergies into more complex actions[53], suggesting that its recruitment here reflects the encoding of objects in terms of their componential motor synergies – an interpretation also supported by the RSA reported below.

The divergence between passive and target object localization is consistent with prior work implicating occipito-temporal cortex in object recognition[4,8,9] and somatosensory and motor regions in the processing of manipulation targets[51,52]. This initial dissociation between the cortical loci of passive and target object variance provides a first indication of remapping of object representations according to contextual role.

An initially unexpected observation was that hand synergy weights generally did not outperform action labels in hand somatomotor brain regions, but rather in occipito-temporal structures, and particularly ventral occipito-temporal regions that support analysis of surface texture and object material properties[14,15,54]. In particular, hand synergy weights explained variance most strongly in a primarily left-lateralized occipital network, including the left superior occipital gyrus, left lingual gyrus, and left cuneus, with additional bilateral contributions in middle occipital gyrus and bilateral fusiform gyrus. Lateral occipital regions are involved in representing body-part configurations[55–57], while ventral occipital regions such as the lingual gyrus likely process surface texture properties of objects related to grip and hand posture. The left lateralization is consistent with a left-hemisphere bias for processing object properties relevant to hand configurations in right-handed individuals[58]. A notable divergence between action labels and hand synergy weights was in the superior parietal lobule, which ranked among the top regions for action labels but was absent for hand synergy features. This outcome is consistent with the role of posterior parietal cortex in encoding goal-directed action representations[24] rather than the fine-grained parameters of motor execution. Regardless, for present purposes, we were interested in isolating the passive and target objects from contributions of action labels and hand synergies—rather than partitioning the variance between action labels and hand postures.

Note that while the winner-take-all map (**Figure 3a**) highlights the dominant feature space at each voxel, most voxels were explained to some extent by all feature spaces. This is apparent across anatomical ROIs (**Figure 3b**), which reveal the relative contribution of each feature space within each region. **Figure 3c** shows that even within winner voxels, for which by definition, a single feature space unambiguously dominates, the remaining feature spaces each contribute additional explained variance (see Supplemental **Figure S3** for whole-brain $R^2$ maps of individual feature spaces).

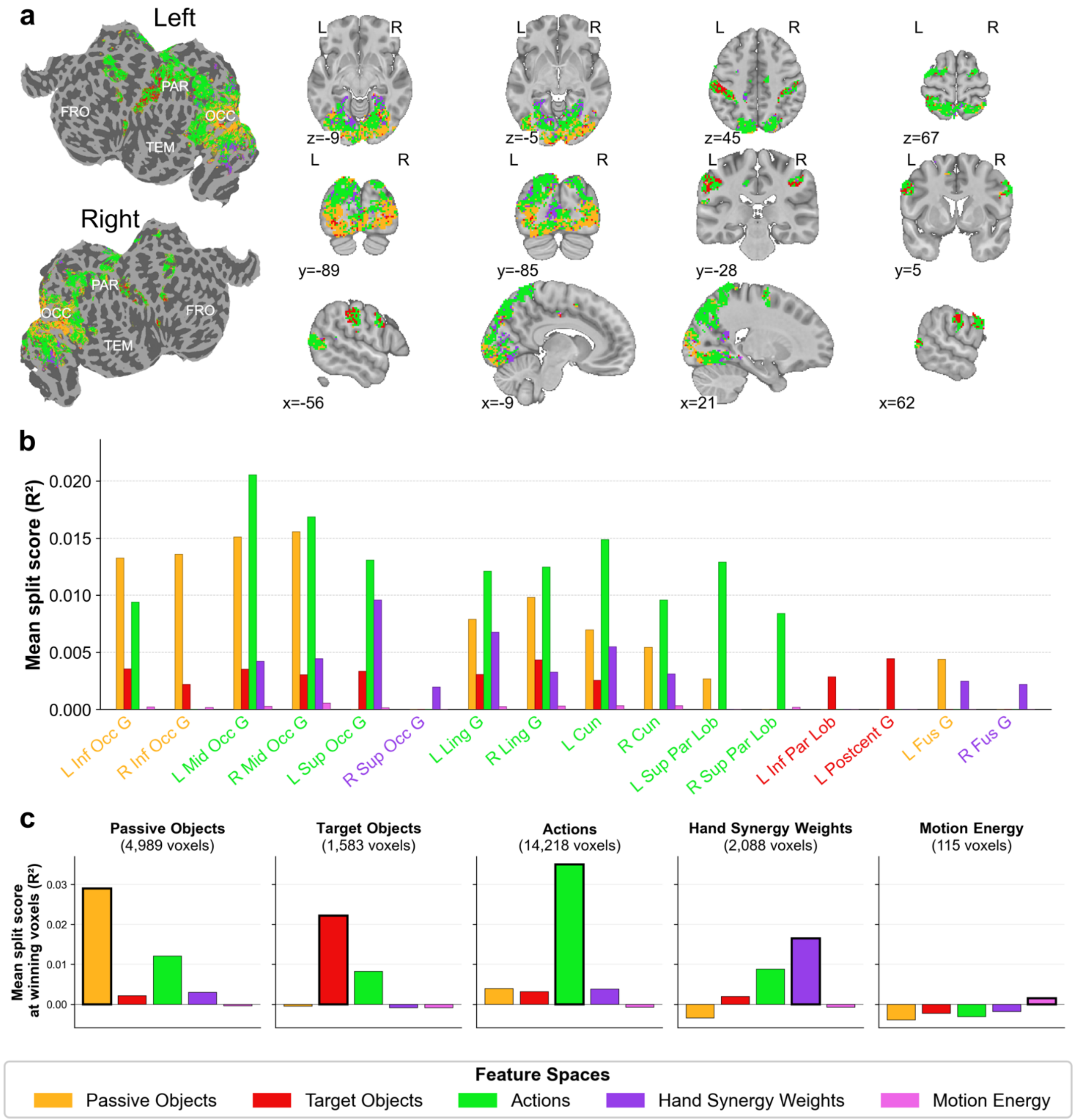

**Figure 3. Whole-brain map of feature space contributions to held-out neural prediction. (a)** Winner-take-all map of feature space contribution to the voxel-wise encoding model. Colors indicate the feature space with the highest mean split score at each voxel. Voxels not surviving group-level FDR correction ($q < 0.05$, minimum cluster extent 10 voxels) are left transparent. Action labels showed the widest cortical coverage, while passive objects localized predominantly to occipito-temporal cortex and target objects to a left-lateralized parietal network including inferior parietal lobule and postcentral gyrus. **(b)** Mean split score ($R^2$) per feature space across ROIs defined as the union of the top 10 regions per feature space, ranked by mean split score. ROI labels are colored according to the feature space with the highest mean split score in that region. Each region showed a dominant feature space alongside contributions from the remaining feature spaces. **(c)** Visualization of mean split score ($R^2$) of all five feature spaces, computed within voxels where each feature space achieved the highest split score, illustrating that while a single feature space dominates

within its winning voxels, other feature spaces can account for additional variance in those same regions. The number of top-predicted voxels for each feature space is shown in parentheses.

### *Contextual role modulates how sensorimotor and occipito-temporal regions represent objects*

To test whether contextual role modulates the engagement of distinct brain networks for object representation, we identified regions differentially representing target versus passive objects by comparing the voxel-wise split scores across contexts ($\Delta R^2$ = target – passive; two-tailed t-test against zero, FDR-corrected $q < 0.05$). This approach allows us to investigate differences in functional recruitment across context. Objects viewed as passive elements of the scene or as action targets showed distinct patterns of preferential object coding, providing evidence for dynamic, context-dependent representation of objects (**Figure 4**). This pattern was consistent across individual participants (Supplemental **Figure S4**).

A clear pattern emerged where target objects were differentially encoded in the left postcentral gyrus and the left supramarginal gyrus, as well as bilaterally in the precentral gyrus. Preferential encoding of target objects in parieto-premotor regions that support visually directed actions[17] and action-relevant representations[51,52] likely reflects a forward model predicting the somatomotor and other sensory consequences of an action. Prior studies have found that viewing small manipulable objects drives differential activity in that same network compared to a range of visual baselines, such as faces or animals[59–61].

In contrast, passive objects preferentially localized to a widespread network encompassing bilateral inferior and middle occipital gyri, bilateral cuneus, left lingual gyrus, left fusiform gyrus, and right superior parietal lobule. These regions have been implicated in multiple aspects of visual object and spatial processing. Lateral occipital cortex plays an important role in object recognition and shape processing[8], while ventral temporal cortex more broadly supports object knowledge[4], including texture, color, categorization, and shape[9] as well as object weight representation[16]. Posterior parietal regions contribute to egocentric representations of space[17] and retrieval of spatial context[18]. The absence of sensorimotor or supramarginal gyrus engagement for passive objects is significant because those structures are differentially active during *passive* viewing of manipulable objects versus nonmanipulable stimuli (faces, animals, places), an observation interpreted as reflecting the category of the stimulus[60,61]. The current findings indicate it is not the 'category' of the visual stimulus that is determinative of neural representational geometries—since the same objects appear as both targets and passive objects—but rather the contextual role the object occupies within the ongoing scene.

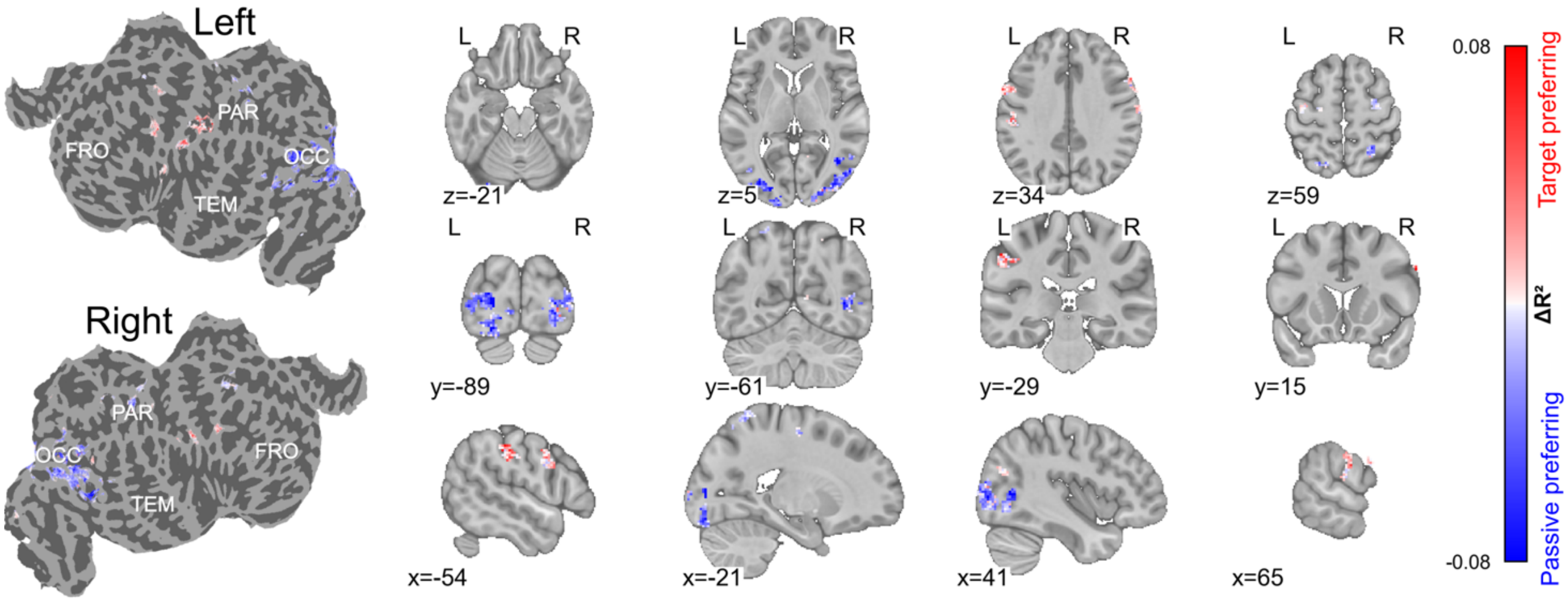

**Figure 4. Context-dependent modulation of object representation across the brain.** Voxel-wise contrast between the split scores of the target object and passive object feature spaces ($\Delta R^2$ = target − passive), revealing regions preferentially predicted by each object context. Contrast map displayed on inflated cortical surfaces and on axial, coronal, and sagittal slices of the MNI template. Red voxels are differentially predicted by target objects; blue voxels are differentially predicted by passive objects. Voxels not surviving group-level FDR correction ($q < 0.05$, minimum cluster extent 10 voxels) are kept transparent. Target objects preferentially engaged a left-lateralized sensorimotor network including postcentral gyrus, inferior parietal lobule, and precentral gyrus, while passive objects preferentially recruited a bilateral occipito-temporal network spanning inferior and middle occipital gyri, lingual gyrus, and fusiform gyrus.

### *Object geometries organize along affordance structure for action target objects and semantic structure for passive objects*

Having identified two distinct networks that preferentially encode objects as passive versus target elements of the scene, we asked how the same objects are represented within each of these context-specific networks. We tested four hypothesis-driven RDMs against neural RDMs constructed within each network: visual similarity (ResNet-50 features[62]), semantic similarity (sentence transformer embeddings; BAAI/bge-base-en-v1.5[63]), action affordances (distribution of actions performed on each object across the video), and hand posture affordances (characteristic hand configurations during manipulation reflecting how an object is typically handled). Networks were defined independently for each participant in subject-native space. Here we were particularly interested in triangulating how context modulates object geometries, while holding constant all other factors. We therefore focused on the set of 26 objects that appeared as both target and passive objects at different points in the videos, ensuring that any observed differences in neural representation could be attributed to contextual role alone, and not to any perceptual properties of the objects themselves.

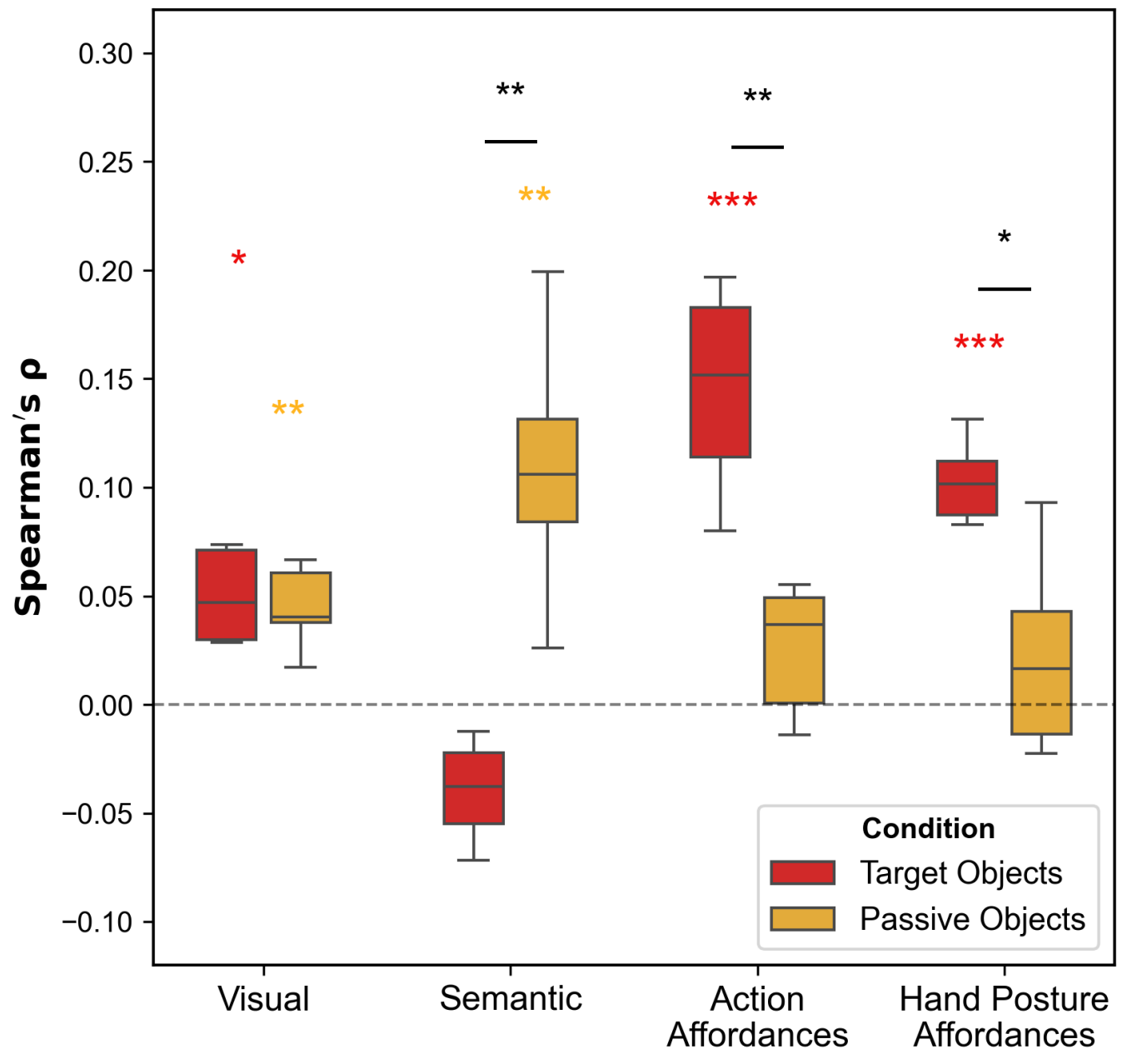

**Figure 5. Spearman correlations between neural and model RDMs for passive and target objects within subject-specific cortical networks.** Results show a double dissociation between matched sets of objects based on context: the neural representational geometry of objects that were action targets aligns with affordance structure (both the actions typically performed with objects and the characteristic hand postures recruited during manipulation), while objects passively viewed in the scene align with semantic structure. Both contexts also show correlation with visual representational structure. Neural RDMs were computed within voxels maximally predicted by target objects and passive objects, respectively, in subject-native space. Colored asterisks indicate significant alignment with each hypothesis RDM within each condition (one-sample t-tests against zero, FDR-corrected); black asterisks with horizontal bars indicate significant differences between conditions (paired t-tests, FDR-corrected). *$p < .05$, **$p < .01$, ***$p < .001$.

Neural representational geometries diverged sharply across the two object contexts (**Figure *5***). When objects were action targets their representations were strongly aligned with action affordances ($t(5) = 7.41$, $p<.001$) and hand posture affordances ($t(5) = 7.36$, $p<.001$) but showed no alignment with semantic similarity ($t(5) = -4.18$, $p = 0.996$). Passive object representations showed the opposite pattern: significant alignment with semantic similarity ($t(5) = 4.53$, $p<.01$) but no reliable alignment with either action affordances ($t(5) = 2.08$, $p = 0.062$) or hand posture affordances ($t(5) = 1.21$, $p = 0.140$). Importantly, regardless of whether objects were action targets or passive elements in the scene, there was comparable alignment with visual similarity (targets: $t(5) = 2.94$, $p<.05$; passive: $t(5) = 4.22$, $p<.01$). Direct paired comparisons confirmed that visual structure was the only dimension that did not differ across context ($p = 0.613$), whereas there was a significant difference between action target and passive contexts for action affordances ($p<.01$), hand posture affordances ($p<.05$), and semantic similarity ($p<.01$; all p-values FDR corrected).

Importantly, visual representational structure was encoded comparably across contexts (**Figure 5**) despite differences in fixation behavior, as was confirmed by a separate eye-tracking experiment described below. This rules out the possibility that differential visual input alone accounts for representational dissociations between contexts.

These findings demonstrate that both target and passive objects maintain visual representational structure in their respective distributed networks but encode fundamentally different types of information beyond visual appearance. Target objects integrate action-relevant properties, specifically the actions typically performed with the object and the characteristic hand configurations used in their manipulation, in line with the involvement of action-relevant regions for targets observed in the localization contrasts (**Figure *4***). In contrast, passive objects encode semantic information reflecting individual object identity, likely supporting an understanding of scene layout and context. This is in line with the widespread visual-spatial network preferentially tuned to passive objects in the localization contrasts. Those regions included lateral occipital cortex and ventral temporal regions implicated in visual object processing, categorization, and recognition.

### *Context-dependent spatial organization of object representational geometries*

To test the anatomic specificity of the above RSA analyses, and to explore how contextual role modulates representational geometries beyond context-selective networks, we conducted a group-level searchlight RSA (**Figure *6***; Supplemental **figures S6–S7** for additional slices).

The results confirmed the anatomical specificity of each representational dimension. For target objects, visual structure drove bilateral lateral occipito-temporal regions, including lateral occipital cortex, which has been shown to support visual shape analysis[8]. For passive objects, visual structure localized across bilateral ventral occipito-temporal regions and dorsal occipital and posterior parietal regions. These regions have been implicated in spatial processing and visuomotor transformations[8,9,58], in line with the interpretation that passive objects are potential obstacles in the environment, and potential next targets of actions.

For target objects, action affordance structure was represented by the left supramarginal gyrus (SMG), as well as the postcentral gyrus. The SMG has been implicated in object manipulation and praxis, and when damaged, can cause upper limb apraxia[27,27,28,31]. Prior work has argued that this area supports a forward model of how object-directed actions will feel—and loss of that mechanism could explain some aspects of upper limb apraxia[27,28,53]. By contrast, for passive objects, action affordance structure was represented along dorsal occipital and IPS regions, including the anterior IPS, which supports visuomotor transformations for hand shaping during object grasping.

Hand posture structure for targets showed a more restricted and predominantly left-lateralized localization, involving the left lingual gyrus, left pre- and post-central gyri. For passive objects, hand posture affordance alignment was predominantly left-lateralized, encompassing occipital cortex, cuneus, fusiform gyrus, middle temporal gyrus, and bilateral lingual gyrus, with weaker additional alignment in left postcentral gyrus.

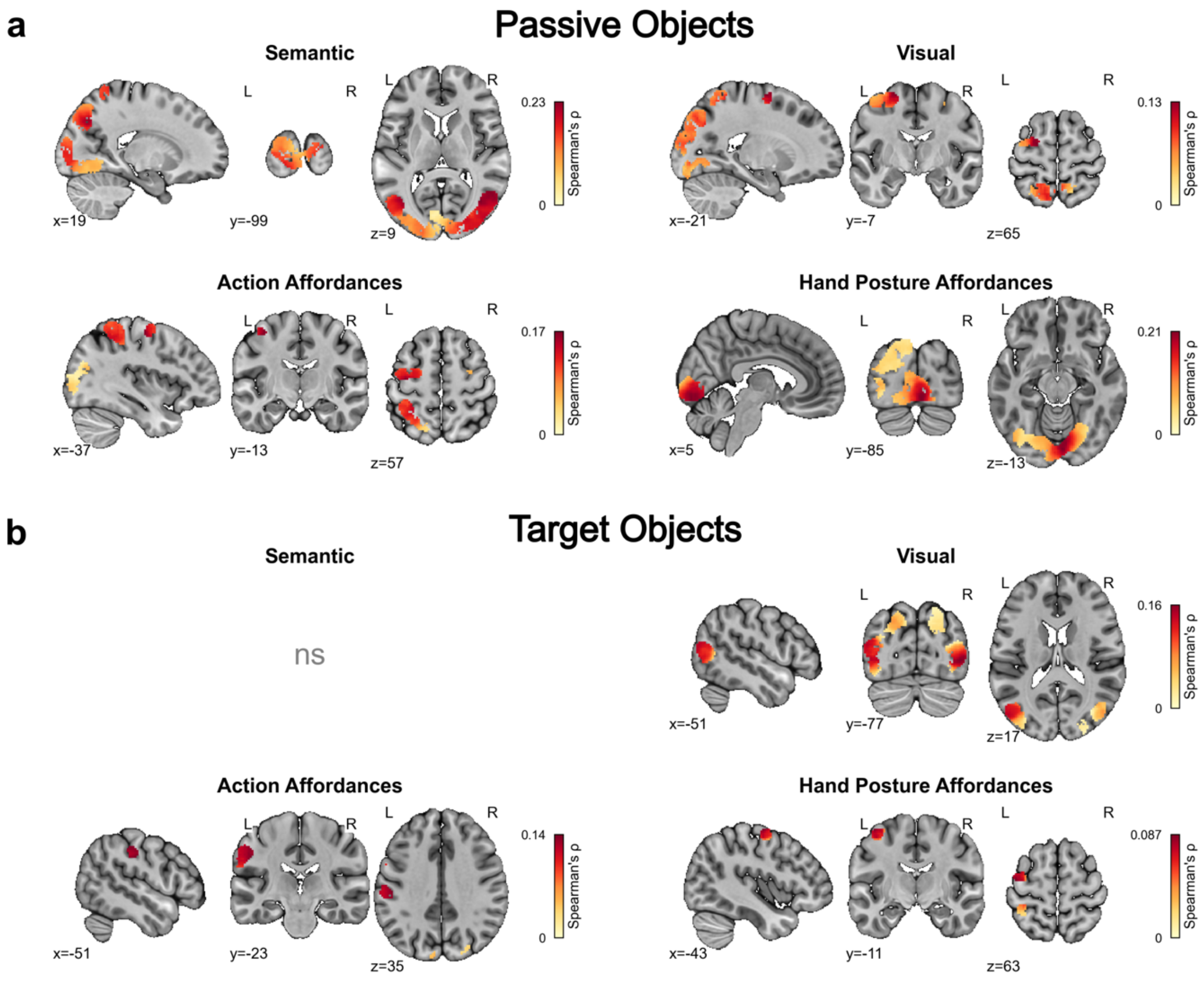

**Figure 6. Searchlight RSA of model RDMs for the same objects when they were action targets and when they were passive elements in the scene.** Searchlight representational similarity analysis (RSA) identified regions whose local representational geometry aligns with each hypothesis RDM, separately for target and passive objects. **(a, b)** Peak slices are shown for each hypothesis and object context. Voxels not surviving group-level FDR correction ($q < 0.05$, minimum cluster extent 10 voxels) are kept transparent; conditions in which no voxels survived correction are indicated as n.s. Additional slices are shown in Supplemental Figures S6–S7.

The presence of affordance-related representational geometry for passive objects in the searchlight, despite its absence in the within-network analysis, is itself informative. Those affordance representations localize outside the voxels most strongly encoding passive objects in the within-subject encoding models and partially overlap with dorsal visual pathway regions along IPS, as well as ventral premotor regions. This suggests that context modulates the degree of engagement of these regions rather than the representational format they employ.

Notably, for target objects, there were no voxels identified for which responses were predicted by the semantic similarity RDM, while for passive objects there was a strong and spatially widespread alignment with the semantic RDM. In fact, semantic structure for passive objects yielded the largest effect sizes of any hypothesis tested, peaking in bilateral middle and inferior occipital gyri and extending into bilateral lingual gyrus, left superior occipital gyrus, left cuneus, and left fusiform gyrus.

Together, the searchlight results extend and spatially constrain the representational dissociation observed in subject-specific analyses. Visual structure is expressed in occipital cortex for objects, both when they are action targets and passive elements in the scene. For target objects, action and hand posture affordance structure extend into somatosensory, premotor, and inferior parietal regions, reflecting the engagement of action planning and somato-motor modeling of objects as action targets. For passive objects, semantic structure dominates with effect sizes that substantially exceed all other hypotheses across both contexts.

***Overt visual attention is preferentially directed toward action targets during naturalistic viewing***

The cortical and representational dissociations reported above raise the question of how overt attentional mechanisms may contribute to the observed modulation. Given that action targets are central to the ongoing scene, we hypothesized that they would capture visual attention more than objects passively present. To investigate this possibility, we ran a separate eye-tracking experiment, which confirmed that action targets were more frequently fixated compared to passive objects (mean fixation probability: targets = 0.20, passive = 0.08; $t(25) = 6.40$, $p < .001$; Supplemental Figure S8), indicating that action targets were more frequently the locus of overt visual attention during naturalistic viewing. Notably, despite this fixation difference, visual representational structure was encoded significantly for both contexts (Passive: $p<.01$; Targets $p<=05$) with no significant difference across contexts ($p = 0.613$) (**Figure *5***), indicating that differential visual input alone does not account for the representational distinction between active and passive contexts.

## DISCUSSION

The neural representation of objects must integrate invariance across input modalities[1,2,64] with flexibility to align stable representations with specific task demands and behavioral contexts[2]. Fundamental questions remain about which dimensions of an object's representation are flexibly modulated by action context and which remain invariant, how each is achieved, and the neural systems involved. Here we leveraged the ecologically rich structure of naturalistic fMRI in which participants viewed first-person cooking videos, keeping object stimuli constant while the movie naturally varied whether those objects were targets of concurrently unfolding actions or passive elements in the display. This goes beyond prior stimulus-based approaches that modeled neural representations by varying stimulus type rather than contextual role.

First, we found that the event structure of actions, over and above objects as either active or passive elements in the scene, was the strongest predictor of neural responses during naturalistic action observation. This establishes that the encoding model captured the dominant structure of the scene, against which the object-specific effects below emerge. Second, contextual role reorganizes the geometry of object representations, producing a double dissociation across the two roles. Action targets engaged a somatomotor network organized by affordance structure, reflecting both the actions an object typically supports, and the characteristic hand configurations used to manipulate it. Passive objects recruited a distributed occipito-temporal network organized along semantic, and not action-relevant, dimensions. This is in contrast with the notion of affordance structure popularized by Gibson[65] which emphasized that the volumetric structure of objects compels action in the absence of a semantic analysis. These findings indicate that the system supports ‘on demand’ remapping of object geometries via dimensions that code the concurrently unfolding action, and in a way that is not merely driven by the visual structure of the images in view. Third, and in contrast, the visual representations of objects remained invariant to context and localized to occipito-temporal regions both when objects were action targets and when they were passive elements in the scene. This reflects an important insight into the ventral stream’s role in supporting a stable representation of an object even as its role in the currently unfolding action changes.

Our findings show that the human brain spontaneously and flexibly promotes specific object geometries, on demand, and based on unfolding scene dynamics—and yet it retains a stable and context-invariant representation throughout. The key finding in this regard is that action context not only leads to differential engagement of action-relevant regions for object processing but reorganizes the representational geometry through which objects are encoded on a changing, moment-to-moment basis. This constitutes a shift in representational content across contexts, not merely a quantitative scaling of response amplitude. When objects are passive elements in the scene, they align most strongly with semantic dimensions in occipito-temporal regions that support object recognition. When objects become the targets of actions, they align strongly with action affordance and hand posture affordance structure, supported by parietal regions known to support skilled object manipulation. The semantic embeddings used in the RSA capture broad distributional similarity reflecting object semantics in a way that is not normative to any particular goal or action context. Affordance structure and hand postures are, by contrast, inherently goal-oriented, reflecting the intersection of an object's properties with a particular action context. Because the core analyses were within the same set of objects, the dissociation between passive and target object representations reflects a shift from a broad, goal-neutral representation of object identity to a goal-oriented representation organized around the actions the object currently supports, as objects dynamically shift between contextual roles. Importantly, because object coefficients were estimated jointly with frame-by-frame action and hand configuration features, this alignment reflects affordance structure carried by object representations themselves, while minimizing the influence of co-occurring motor variables.

These findings illustrate how flexibility and invariance can coexist within the same representational system. Visual representational structure was encoded in a context-invariant way, suggesting that the visual properties of objects are encoded regardless of contextual role, and constitute a stable representational format across contexts. This supports the view that objects are represented in a stable, invariant manner while also allowing for flexibility over representational dimensions to support context-specific demands. Consistent with this account, prior work suggests that visual features can contribute to stimulus-invariant representations[1] and that early visual regions are more task-independent than higher-order regions, which show substantial task-dependent modulation[37]. The searchlight analysis further supports the view that contextual role modulates the degree to which different representational dimensions dominate, rather than determining which dimensions are present in an absolute sense: affordance-related representational geometry remains detectable for passive objects in action-relevant parietal regions, even though it was absent in the networks most strongly encoding passive objects. This reflects a change in the degree to which regions encoding context-relevant information are recruited, rather than a change in the nature of information that is encoded in a given region.

Together, these results speak to a broader question about whether representational flexibility and invariance are mutually exclusive hypotheses about the operation of the semantic system[66,67]. One perspective is that flexibility in behavior indicates less structure in the system, which is to say a lesser role for invariant representations. Another perspective is that flexibility is a consequence of the structure of the system. Our findings indicate that flexibility and invariance are achieved at different levels, and over separable types of information, during dynamic scene processing.

What mechanisms support context-dependent representational modulation? Modulation of representational geometries can be thought of as a form of attentional modulation of conceptual representations during naturalistic viewing. In a cooking scene, action targets are the objects central to the ongoing events and therefore essential to understanding the unfolding scene. Prior work has shown that attentional mechanisms emphasize target-relevant information[68,69], including evidence that selective attention to a behaviorally relevant category shifts voxel-wise semantic representations toward the attended target during naturalistic viewing[48]. Computational and single-unit evidence suggests that such attention-driven tuning shifts reflect a matched-filter mechanism that expands the cortical representation of task-relevant features at the expense of others[70,71]. In the context of the present study, attention to the action targets promotes their action-affordance properties in their neural representation while dampening the representation of goal-neutral semantic features that are less relevant to interpreting the ongoing action. This attentional account is supported by the eye-tracking results, showing overt visual attention was preferentially directed toward action targets. Notably, despite this fixation difference, visual representational structure was encoded comparably across both contexts, arguing against the possibility that differential visual input alone accounts for the modulation of representations across contexts.

A complementary mechanism through which such representational modulation may be implemented is the semantic control network. This framework proposes that context-relevant representations are supported by a control system anchored in left inferior frontal cortex and posterior middle temporal gyrus, which modulates activation within a distributed representational system toward task-relevant dimensions of an object[2,72]. Under this view, recognizing an object as the current target of a goal-directed action may recruit this network to selectively upweight affordance-relevant features at the expense of goal-neutral semantic content, producing downstream representational reorganization in the occipital, fusiform, and somatosensory regions sampled by our analyses. Consistent with this, hierarchical control frameworks propose that mid-lateral prefrontal cortex exerts top-down influence over premotor and sensory-motor regions[73], which broadly overlap with the target object network identified here, suggesting that frontal control signals may drive the representational reorganization observed in these posterior regions.

The present results establish that representational modulation can occur in the absence of any explicit task instruction, arising instead from the functional relevance of the objects within the naturalistic scene itself. This complements prior evidence from explicit tasks and points toward a more general principle: the functional relevance of an object within an ongoing behavioral context is sufficient to modulate its representational geometry in the absence of explicit top-down control. Future research with higher temporal resolution can test the dynamics of flexible modulation of conceptual content. In particular, electrophysiological methods, including magnetoencephalography and intracranial recordings, offer the temporal resolution needed to map when, in the flow of processing, the representational geometry of an object is reshaped by its behavioral context, and how rapidly this modulation occurs as contextual role dynamically shifts in the ongoing scene. More broadly, a full characterization of invariance and flexibility must go beyond conceptual retrieval. For practical reasons, most empirical work measures representations once they are already deployed under task or contextual influence, meaning invariance is operationalized as stability across retrieval instances. But this may not capture the format of stored object concepts prior to their retrieval. Understanding what might be referred to as 'dark content' — or representational content prior to engagement and retrieval — may be crucial for a complete account of invariance and flexibility.

## ONLINE METHODS

### *Subjects*

Eight participants consented to take part in the experiment. Two participants were excluded for failing to pass quality assurance tests (see "fMRI preprocessing"). The remaining subjects had a mean age of 26.86±2.48 (5 female, 1 male), were right-handed, healthy and had normal or corrected-to-normal vision. The study receive approval from the Carnegie Mellon University's Institutional Review Board.

### *Movie stimuli and annotations*

The stimuli consisted of a movie shot in first-person view, depicting a cooking scene in which a person prepares food in a kitchen, performing actions (e.g. grasping, cutting, scrubbing) on a variety of objects (e.g., onion, knife, fridge). The movie was compiled using a subset of the Epic-Kitchens dataset [74,75]. The selected clips were chosen in a way that attempts to maximize the incidence of actions with multiple objects using the original Epic-Kitchens' metadata. Qualitative choices were then made to minimize visual clutter in the scenes (e.g. crowded kitchen counter, excess of unused objects scattered in the scene). Final trims from the resulting clips were also made to minimize times of idleness in the scene (e.g. multiple minutes waiting for a dish to warm up). The final movie was 46 minutes and 50 seconds long (sampling frequency of 24Hz) and was divided into 5 runs of approximate equal lengths (run 1:09:19, runs 3-4: 09:20, run5: 09:31).

Movie annotations were done on a down-sampled version of the stimuli at 4Hz (11232 frames). Initial annotations were produced by trained human annotators through the CVAT.ai annotation service [76]. We provided detailed instructions to the annotators, including comprehensive examples of properly annotated frames. The annotation protocol specified precise criteria for identifying and labeling relevant objects and actions within each frame. Each frame was annotated for all visible objects using bounding boxes, the ongoing action, the target object of the action, and the tool used (if any). The annotation process involved three iterative rounds of revision. In each round, the annotators provided all the annotated frames which we reviewed comprehensively and compiled detailed feedback and adjustment requests that were sent back for refinement. Inter-annotator reliability was assessed on 1,120 frames (~10% of the dataset) that were independently labeled twice during the initial annotation phase, yielding high agreement for object identification and for action labeling (respectively 92.8% and 78.3% mean Jaccard similarity). Following the final annotation phase, trained members of the research team conducted an additional frame-by-frame manual review to ensure annotation quality and consistency, with particular attention to systematically refining action labels. Any discrepancies or ambiguities were resolved through discussion and standardized according to our predetermined labeling criteria.

Additional movie features were extracted using computational vision methods. Hand skeletal data were obtained using Mediapipe Hands[77] which identified 21 3D landmarks for each hand across all frames. Hand postures were then mapped onto a set of 20 hand kinematic synergies derived from Caglar et al. (2025)[53], representing minimal kinematic units that combine to form complex object-directed actions. The final frame of each reference GIF provided by Caglar et al. (2025) was used as the canonical representation of each synergy (see Figure S8). The mapping procedure involved Procrustes alignment to achieve rotational and translational invariance across hand configurations. All hand configurations (from both the 20 reference synergies and movie frames) were then normalized by centering and scaling them to lie within the unit sphere. An iterative alignment procedure was then applied: at each iteration, hand configurations were aligned to a mean reference configuration using a rotation matrix that minimized alignment error between each hand configuration and the mean reference configuration, and the mean configuration was recomputed from the aligned configurations. This process converged when the change in the mean configuration fell below 1e-6. Following alignment, each hand configuration was rotated to align its principal axis with the z-axis to establish a canonical orientation. A vector of weights was computed for each frame by solving a least-squares problem where movie hand configurations were expressed as linear combinations of the 20 reference synergies. Each weight quantified the contribution of its corresponding reference synergy to the observed hand posture at each frame and served as regressors in the fMRI analysis. This remapping approach was chosen rather than using raw hand skeletal landmarks because the synergy weights provide rotation- and translation-invariant representations of hand configuration, making postures comparable across different spatial positions in the video. Additionally, the synergies capture compositional, action-related hand configurations rather than arbitrary joint positions and reduce the number of free parameters in the encoding model.

Motion energy features were extracted as nuisance covariates to account for low-level visual motion. Frames were first extracted from the original 24 Hz movie stimuli and converted to grayscale luminance images. Each frame was resized to 96×96 pixels and scaled to a 0–100 luminance range. Motion energy was computed using a multi-scale pyramid with spatial frequencies of 0, 2, 4, 8, 16, and 32 cycles per image, implemented via the pymoten package[78]. The pyramid decomposed each frame into spatiotemporal filters at multiple scales and orientations, capturing local motion patterns. Features were initially computed at 24 Hz, then downsampled to 4 Hz using finite impulse response filtering with zero-phase delay to match the temporal resolution of the manual annotations. Finally, all motion energy features were averaged across filters and orientations to obtain a single summary measure per frame.

### *Experimental procedure*

All subjects participated in two identical sessions of the experiment. During each session all runs were presented once except the third run which was presented an additional time at the end of the experiment and later served as the validation run for the voxel-wise encoding models. This repeated-run design was intended to improve signal-to-noise in the within-subject encoding model estimates. Each run started with a fixation cross that remained on screen for 20 seconds, followed by the presentation of the movie, and finally another fixation cross for 20 seconds. Stimuli were displayed on a screen with 1512×982 pixel resolution positioned at 55 cm from the subject, with the movie occupying a centered region of 1472×828 pixels. During each session, subjects signed a consent form, underwent fMRI scanning, and then completed a debrief experiment aimed at assessing their attention during the task. Participants were instructed to watch the movie and remained as engaged as possible with the events happening in the scene.

The debrief experiment consisted of two tasks. The first task showed a subset of the frames from the movie along with other frames that did not appear in the movie and were taken from other scenes of the EpicKitchens dataset. Participants were instructed to press one of two buttons to indicate whether the frame appeared in the movie they watched or not. The second task presented pairs of frames from the movie stimuli and asked participants to indicate using two buttons which of the two frames occurred later in the movie relative to the other one. Mean accuracies were high, indicating that subjects were attentive during the scanning sessions (0.97 ± 0.02 for the first task and 0.88 ± 0.14 for the second task).

### *fMRI data collection*

MRI data were acquired on a 3T Siemens Prisma scanner (Siemens Healthcare, Erlangen, Germany) at the CMU-Pitt Brain Imaging Data Generation & Education (BRIDGE) Center using a 64-channel Head/Neck coil. Structural images were collected using a T1-weighted 3D MPRAGE sequence (TR = 2300 ms, TE = 1.99 ms, inversion time = 900 ms, flip angle = 9°). The acquisition had a 1 mm isotropic resolution (slice thickness = 1 mm), 256 × 256 base resolution, 100% phase FOV, and GRAPPA acceleration (in-plane factor = 2). Images were acquired with a gradient-echo inversion-recovery sequence (tfl3d1_16ns) with adaptive coil combination and no nonlinear gradient correction applied. Functional images were acquired using a 2D multiband echo-planar imaging (EPI) sequence (cmrr_mbep2d_bold) with multiband factor 3. Data were collected with TR = 2000 ms, TE = 30 ms, flip angle = 79°, slice thickness = 2 mm with 2 mm spacing, and 106x106 in-plane base resolution. Partial Fourier (6/8) and 100% phase FOV were used. The sequence incorporated PFP, fat-saturation, and extended dynamic range options, with a total readout time of 69.3 ms and effective echo spacing of 0.66 ms. Phase-encoding was performed in the j- direction.

### *fMRI preprocessing*

Functional MRI data were preprocessed using AFNI 25.0.06 (afni_proc.py) [79,80]. The first two TRs were removed from each run. Preprocessing included slice timing correction, alignment of functional images to the anatomical image using local Pearson correlation with Z-score transformation (lpc+ZZ cost function), spatial normalization to MNI152_2009 template space, motion correction via volume registration to the volume with minimum outliers, masking, and intensity scaling. Motion correction included alignment of functional images to the anatomical image and concatenation with the nonlinear warp to template space. Additional processing steps were implemented using custom scripts: Data were linearly detrended and high-pass filtered via median subtraction with a 120 s window. Time series were z-scored along the time axis. For visualization plots on the flattened surface, surface reconstruction was performed using FreeSurfer's recon-all (version 7.4.1)[81] on the processed anatomical images. The resulting surfaces were imported into Pycortex 1.2.11[82] for the visualization over flat maps. A cortical mask was generated for each subject using Pycortex (type='cortical'). All subsequent analyses are restricted to the voxels contained within this mask.

Data quality was assessed by comparing voxel-wise correlations between repeated runs versus unrelated runs over the time dimension. Since run 3 was repeated as run 6, these runs presented identical stimuli and served as the repeated condition. For each participant and session, voxel-wise correlations were computed between all pairwise combinations of runs. Participants were excluded if the mean correlation of the repeated runs was lower than the mean correlation of all unrelated run pairs. Two participants were excluded, and post-scan debriefing indicated these participants reported drowsiness or falling asleep during scanning.

### *Voxel-wise encoding models*

**Feature preprocessing.** The input features to the encoding model consisted of the human annotations of the movie stimuli as well as model-based annotations of hand postures and motion energy. For the hand posture primitive weights, a binary indicator variable was added for each hand to distinguish frames where the hand was visible on screen from those where it was absent. Hand posture weights and motion energy features were z-scored over the time dimension within each run. Human-annotated action features were one-hot encoded, while passive objects and target objects were multi-hot encoded to allow simultaneous object representations when multiple objects are involved in a contextual role. Features occurring in fewer than 50 TRs across the training runs were excluded from analyses.

**Model specifications.** Repeated runs were averaged within and across sessions. Test runs 3 and 6 (each with 4 repeats) were averaged to form a single held-out test run. Training runs 1, 2, 4, and 5 (each with 2 repeats) were similarly averaged, yielding four training runs. Movie features were temporally aligned with the fMRI data using Lanczos downsampling (window = 3, cutoff multiplier = 1.0)[83]. fMRI responses were modeled using voxel-wise encoding models with finite impulse response (FIR) delays to capture the hemodynamic response, using 4 delays of 1-4 TRs (2, 4, 6, and 8 seconds). Movie features were defined as belonging to five different feature spaces (Passive objects, target objects, actions, hand synergy weights, and motion energy).

Models were fit using banded ridge regression implemented in the *Himalaya* Python package (version 0.4.6)[50], which applies separate regularization parameters to different feature spaces optimized via cross-validation. This approach accounts for potential dependencies among features within each space while allowing independent regularization across spaces, enabling us to quantify the relative contribution of each feature space to model predictions. Regularization parameters were selected from 20 logarithmically spaced values (ranging from $10^1$ to $10^{20}$) using random search with 40 iterations. Cross-validation followed a nested structure. Hyperparameter optimization was performed using leave-one-run-out cross-validation within the training set, where models were iteratively trained on three runs and validated on the fourth. Model performance was evaluated using voxel-wise $R^2$ between predicted and observed BOLD responses on the held-out test run.

**Statistical testing of model performance.** Group-level significance was assessed in voxels modeled in at least 3 subjects. Individual subject $R^2$ maps were smoothed with a 6mm FWHM

Gaussian kernel and then group statistics were computed using voxel-wise one-tailed one-sample t-tests testing $R^2$ scores against zero. Multiple comparisons were FDR-corrected ($q < 0.05$) with a minimum cluster extent of 10 voxels.

We also assessed the statistical significance of each subject's model performance. We used a voxel-wise permutation test with 1,000 iterations. The test evaluated whether predicted voxel time series were significantly correlated with observed BOLD responses by generating a null distribution through temporal permutation of the predictions. To preserve temporal autocorrelation structure in the fMRI data, permutations were applied in chunks of 10 TRs (20 seconds), where consecutive TRs were kept together during shuffling. Empirical p-values for each voxel were computed as the proportion of permuted correlations that exceeded the observed correlation. Multiple comparisons across voxels were corrected using FDR at $\alpha = 0.05$.

All post-hoc analyses were restricted to voxels that showed significant model scores at the group level, except for the RSA at top predicted voxels, which was performed using subject-specific significance maps. The group-level significance maps ensured spatial correspondence across subjects for group-level localization results, while the subject-specific approach focused on preserving individual-specific functional organization in the RSA at top predicted voxels.

**Quantifying feature-space contributions.** To quantify how each feature space contributed to voxel-wise predictions, we decomposed the explained variance of each subject's joint model using the product-measure split score (r2_score_split from *Himalaya*). The product-measure decomposition partitions the joint $R^2$ into additive components for each feature space while accounting for correlations among their predicted time courses. This ensures that the component scores sum exactly to the joint model $R^2$. Split scores were computed at group-level significant voxels using individual subject $R^2$ maps. Results are plotted as a winner-takes-all map indicating the feature space with the highest split score at each voxel. Results are also summarized in a bar graph showing the mean split score for each feature space across a pooled set of ROIs, defined as the union of the top 10 ROIs for each feature space, with anatomical ROIs defined using the Talairach Atlas labels[84,85].

**Coefficient extraction.** To examine feature-specific tuning patterns, we extracted regression coefficients from the fitted banded ridge models. Because banded ridge regression optimizes regularization parameters independently per voxel, coefficients can have very different scales. To correct for this, coefficients were rescaled to have a norm equal to the square root of each voxel's $R^2$ score[86]. This rescaling ensures that voxel-wise coefficients are on a comparable scale when computing pairwise similarities within a single feature space across voxels and does not affect similarity computations across feature spaces. Normalized coefficients were then averaged across FIR delays to obtain a single weight per feature.

### *Targeted contrasts*

To investigate how functional recruitment of brain regions is modulated by the contextual role of objects, we compared voxel-wise explained variance of objects when they appeared as passive objects in the scene versus as targets of actions. For each subject, a contrast comparing the product-measure split score ($R^2$) of target objects and passive objects was computed (targets minus passives). Group-level significance was assessed by smoothing individual subjects' contrast maps with a 6mm FWHM Gaussian kernel and then computing a voxel-wise two-tailed one-sample t-tests testing the contrast against zero. Multiple comparisons were FDR-corrected ($q < 0.05$) with a minimum cluster extent of 10 voxels. Results are plotted as whole-brain voxel-wise maps and summarized in bar graphs showing the 10 ROIs with the strongest average positive modulation and the 10 ROIs with the strongest average negative modulation, with ROIs defined using the Talairach Atlas labels[84,85].

***Representational Similarity Analyses (RSA)***

Beyond identifying which voxels show context-dependent tuning, we investigated how the representational structure of individual objects is modulated by their context in the videos. We focused on the 26 objects that appeared in both contexts (74.29% of objects; total=35), enabling direct comparison across contexts for a matched set of items. We employed representational similarity analysis (RSA)[87] to compare the representational geometry of passive and target objects as captured by the encoding model against the representational geometry predicted by four hypotheses described below. All primary RSAs were conducted on the 26 objects that appeared in both passive and target contexts, except for the hand posture affordances hypothesis where three objects (cupboard, oven/stove, sauce) were excluded because they were never directly manipulated by the hands, resulting in 23 objects for this hypothesis.

**Hypothesis RDMs.** Each RDM was computed as pairwise correlation distances (1 - Pearson r) between embeddings of the corresponding models. Hypothesis RDMs were largely independent, with a maximum pairwise correlation of $r=0.13$ between the visual and semantic RDMs. The model embeddings are described below:

- *Semantic:* Semantic embeddings were created for each object using a pretrained sentence transformer model (BAAI/bge-base-en-v1.5)[63]. A sentence transformer was used to account for labels that had more than one word (i.e. pizza box, oven mitts). For the “carton/canister” and “sponge/rag” labels the embedding was computed for each of the two alternative words and averaged for each pair. All embeddings were L2-normalized.
- *Visual:* Visual embeddings were extracted using a pretrained ResNet-50 model[62] (ImageNet1K_V1 weights from torchvision version 0.20.1). For each object, all instances across the movie annotations were identified using bounding boxes from the human annotations. To equalize sampling across objects, we randomly sampled 622 instances per object (where 622 equals the number of annotated instances for the least frequent object). Each bounding box region was cropped from the corresponding movie frame, resized and normalized, and passed through ResNet-50. Features from the average pooling layer (2048-dimensional) were extracted and L2-normalized for each instance. The final embedding for each object was computed as the mean across sampled instances, then renormalized.
- *Action affordances:* Action affordances captured the distribution of actions performed with each object. For each exemplar, we extracted all frames from the human annotations in which the object appeared during an action. Action co-occurrence vectors were constructed by counting how many times each unique action appeared with each object. These count vectors were L2-normalized to create affordance embeddings.

- *Hand posture affordances:* Hand posture affordances captured a prototypical hand configuration for each object, along with the variability around this configuration across all instances of manipulation. For each object, we identified all frames where the object was directly manipulated by hand and extracted the corresponding hand posture primitive weights. The mean and standard deviation of these features were concatenated to form a single object-level embedding vector and L2-normalized. Although derived from the same movie annotations used in the encoding models, these embeddings summarize object-level statistics rather than frame-wise voxel activity. Critically, representational similarity is evaluated within target object and passive object coefficients specifically, which reflect variance attributable to object identity after the frame-wise contributions of co-occurring actions and hand configurations have been partitioned into their own feature spaces by the banded ridge regression. Finding that affordance geometry aligns with these object-specific representations therefore would indicate that affordance structure is carried by object identity signals themselves, beyond variance explained by the actions and hand postures that co-occurred with those objects in the frame-by-frame stimulus.

**RSA at subject-specific maximally predicted voxels.** This approach first functionally localizes the brain networks most sensitive to each object context in subject-native space. The RSA then independently tests the alignment of representational geometry of objects within these networks against hypothesis RDMs, and whether this alignment differs across networks. By operating at the level of representational geometry rather than individual voxel responses[87], RSA allows us to directly characterize how representational structure differs across the two object contexts in their spatially distinct networks. To preserve subject-specific functional organization, for each subject, we localized voxels maximally predicted by each feature space (target objects and passive objects) based on the product-measure split scores. For each of the objects, we obtained the averaged encoding model weights across delays for both the target objects and passive objects feature spaces. This yielded two coefficient vectors per object—one representing the object when it appears as a target and one when it appears passively in the scene—where each vector contained coefficients across all voxels maximally predicted by that feature space for that subject. These whole-brain coefficient vectors were then used to construct neural RDMs by computing pairwise correlation distances (1 - Pearson r) between all object pairs, resulting in two neural RDMs per participant. To test whether each hypothesis RDM explained the neural representational geometry, we computed Spearman correlations between each hypothesis RDM and the neural RDMs for both target and passive objects. Statistical significance was assessed using one-sample t-tests (one-tailed, testing against zero) across participants, with FDR correction applied within each condition (target, passive) across the four hypothesis tests. To evaluate whether target objects showed stronger or weaker alignment with each hypothesis compared to passive objects, we additionally performed paired t-tests comparing the correlation strengths between conditions.

**Searchlight RSA.** To examine the spatial extent of representational similarities with the hypothesis RDMs, we also conducted a searchlight representational similarity analysis. Each searchlight consisted of a sphere with radius of 5 voxels (10 mm) centered on voxels that showed significant encoding model performance at the group level. Searchlights were retained only if at least 50% of voxels within the sphere were significantly predicted by the encoding models, ensuring analyses focused on well-modeled cortical regions. For each searchlight, neural representational dissimilarity matrices (RDMs) were constructed by computing pairwise correlation distances (1 - Pearson ρ) between object coefficient patterns within the sphere. This procedure was performed separately for target objects and passive objects, yielding neural RDMs that captured how objects were represented in local cortical patches under each condition. Each hypothesis RDM was compared against the neural RDMs using Spearman rank correlation, implemented in the rsatoolbox Python package[88]. Individual subjects' searchlight correlation maps were Fisher z-transformed (arctanh). Group-level significance for each model was assessed using one-sample t-tests against zero on the z-transformed maps, following the same statistical procedures as the encoding model analyses (6mm FWHM smoothing, voxel-wise FDR correction at $q < 0.05$, minimum cluster extent of 10 voxels). Results are plotted as whole-brain voxel-wise maps. For each model RDM and object context, we also identified the 10 ROIs showing the highest Spearman ρ (Talairach Atlas labels[84,85]) and visualized these as a heatmap ($\log_{10}$ scale) to facilitate cross-model comparison.

### ***Eye-tracking experiment***

A separate eye-tracking experiment was run to assess whether action targets and passive objects differed in fixation rates during naturalistic viewing. Eye-tracking data were collected during fMRI scanning but could not be used due to poor signal quality arising from coil glare and scanner vibration. The eye-tracking experiment was therefore conducted as a separate behavioral experiment with an independent sample of 34 participants (16 female, 13 male, 1 other; mean age 20.6 ± 1.4 years; 82% right-handed), providing a normative characterization of gaze behavior during naturalistic viewing of the movie stimulus. Participants who did not take part in the fMRI study viewed the same movie stimuli, without run repetitions, while gaze was recorded binocularly using a Tobii eye-tracker (sampling rate = 60 Hz; display size 1920 × 1080 pixels, video region centered on screen with size 1472 × 828 pixels). Gaze data were downsampled to 4 Hz to match the annotation frame rate. Fixation was defined as gaze coordinates falling within the annotated bounding box of an object on a given frame. Fixation probability was computed per frame as the proportion of participants meeting this criterion. For each of the 26 objects appearing in both contextual roles, mean fixation probability was computed separately for frames in which the object was a target and for frames in which it was passively present. A paired t-test across exemplars compared fixation probability between the two conditions.

## References

1. Dirani, J. & Pylkkänen, L. MEG evidence that modality-independent conceptual representations contain semantic and visual features. *J. Neurosci.* **44**, (2024).
2. Ralph, M. A. L., Jefferies, E., Patterson, K. & Rogers, T. T. The neural and computational bases of semantic cognition. *Nat. Rev. Neurosci.* **18**, 42–55 (2017).
3. Hebart, M. N., Bankson, B. B., Harel, A., Baker, C. I. & Cichy, R. M. The representational dynamics of task and object processing in humans. *elife* **7**, e32816 (2018).
4. Martin, A. The Representation of Object Concepts in the Brain. *Annu. Rev. Psychol.* **58**, 25–45 (2007).
5. Haxby, J. V. *et al.* A common, high-dimensional model of the representational space in human ventral temporal cortex. *Neuron* **72**, 404–416 (2011).
6. Hebart, M. N. *et al.* THINGS: A database of 1,854 object concepts and more than 26,000 naturalistic object images. *PloS One* **14**, e0223792 (2019).
7. Kriegeskorte, N. *et al.* Matching categorical object representations in inferior temporal cortex of man and monkey. *Neuron* **60**, 1126–1141 (2008).
8. Grill-Spector, K., Kourtzi, Z. & Kanwisher, N. The lateral occipital complex and its role in object recognition. *Vision Res.* **41**, 1409–1422 (2001).
9. Grill-Spector, K. & Weiner, K. S. The functional architecture of the ventral temporal cortex and its role in categorization. *Nat. Rev. Neurosci.* **15**, 536–548 (2014).
10. Mishkin, M., Ungerleider, L. G. & Macko, K. A. Object vision and spatial vision: two cortical pathways. *Trends Neurosci.* **6**, 414–417 (1983).
11. Merigan, W. H. & Maunsell, J. How parallel are the primate visual pathways? *Annu. Rev. Neurosci.* https://psycnet.apa.org/record/1993-32418-001 (1993).
12. Goodale, M. A. & Milner, A. D. Separate visual pathways for perception and action. *Trends Neurosci.* **15**, 20–25 (1992).

13. De Haan, E. H. & Cowey, A. On the usefulness of ‘what’and ‘where’pathways in vision. *Trends Cogn. Sci.* **15**, 460–466 (2011).

14. Cavina-Pratesi, C., Kentridge, R. W., Heywood, C. A. & Milner, A. D. Separate processing of texture and form in the ventral stream: evidence from FMRI and visual agnosia. *Cereb. Cortex* **20**, 433–446 (2010).

15. Simmons, W. K. *et al.* A common neural substrate for perceiving and knowing about color. *Neuropsychologia* **45**, 2802–2810 (2007).

16. Gallivan, J. P., Cant, J. S., Goodale, M. A. & Flanagan, J. R. Representation of object weight in human ventral visual cortex. *Curr. Biol.* **24**, 1866–1873 (2014).

17. Kravitz, D. J., Saleem, K. S., Baker, C. I. & Mishkin, M. A new neural framework for visuospatial processing. *Nat. Rev. Neurosci.* **12**, 217–230 (2011).

18. Burgess, N., Maguire, E. A., Spiers, H. J. & O’Keefe, J. A temporoparietal and prefrontal network for retrieving the spatial context of lifelike events. *Neuroimage* **14**, 439–453 (2001).

19. Rizzolatti, G. & Matelli, M. Two different streams form the dorsal visual system: anatomy and functions. *Exp. Brain Res.* **153**, 146–157 (2003).

20. Binkofski, F. & Buxbaum, L. J. Two action systems in the human brain. *Brain Lang.* **127**, 222–229 (2013).

21. Xu, Y. A Tale of Two Visual Systems: Invariant and Adaptive Visual Information Representations in the Primate Brain. *Annu. Rev. Vis. Sci.* **4**, 311–336 (2018).

22. Kravitz, D. J., Saleem, K. S., Baker, C. I., Ungerleider, L. G. & Mishkin, M. The ventral visual pathway: an expanded neural framework for the processing of object quality. *Trends Cogn. Sci.* **17**, 26–49 (2013).

23. Hickok, G. The role of mirror neurons in speech perception and action word semantics. *Lang. Cogn. Process.* **25**, 749–776 (2010).

24. Culham, J. C. & Valyear, K. F. Human parietal cortex in action. *Curr. Opin. Neurobiol.* **16**, 205–212 (2006).

25. Caspers, S. *et al.* The human inferior parietal cortex: cytoarchitectonic parcellation and interindividual variability. *Neuroimage* **33**, 430–448 (2006).

26. Caspers, S. *et al.* The human inferior parietal lobule in stereotaxic space. *Brain Struct. Funct.* **212**, 481–495 (2008).

27. Buxbaum, L. J., Kyle, K. M. & Menon, R. On beyond mirror neurons: Internal representations subserving imitation and recognition of skilled object-related actions in humans. *Cogn. Brain Res.* **25**, 226–239 (2005).

28. Buxbaum, L. J., Johnson-Frey, S. H. & Bartlett-Williams, M. Deficient internal models for planning hand–object interactions in apraxia. *Neuropsychologia* **43**, 917–929 (2005).

29. Rizzolatti, G. & Craighero, L. THE MIRROR-NEURON SYSTEM. *Annu. Rev. Neurosci.* **27**, 169–192 (2004).

30. Tranel, D., Kemmerer, D., Adolphs, R., Damasio, H. & Damasio, A. R. NEURAL CORRELATES OF CONCEPTUAL KNOWLEDGE FOR ACTIONS. *Cogn. Neuropsychol.* **20**, 409–432 (2003).

31. Mahon, B. Z. & Almeida, J. Reciprocal interactions among parietal and occipito-temporal representations support everyday object-directed actions. *Neuropsychologia* **198**, 108841 (2024).

32. Binder, J. R., Desai, R. H., Graves, W. W. & Conant, L. L. Where is the semantic system? A critical review and meta-analysis of 120 functional neuroimaging studies. *Cereb. Cortex* **19**, 2767–2796 (2009).

33. Gan, G., Büchel, C. & Isel, F. Effect of language task demands on the neural response during lexical access: a functional magnetic resonance imaging study. *Brain Behav.* **3**, 402–416 (2013).

34. Xu, Y. *et al.* Doctor, teacher, and stethoscope: neural representation of different types of semantic relations. *J. Neurosci.* **38**, 3303–3317 (2018).

35. Wang, X. *et al.* Representational similarity analysis reveals task-dependent semantic influence of the visual word form area. *Sci. Rep.* **8**, 3047 (2018).

36. Kiefer, M. Perceptual and semantic sources of category-specific effects: Event-related potentials during picture and word categorization. *Mem. Cognit.* **29**, 100–116 (2001).

37. Harel, A., Kravitz, D. J. & Baker, C. I. Task context impacts visual object processing differentially across the cortex. *Proc. Natl. Acad. Sci.* **111**, (2014).

38. Bar, M. & Ullman, S. Spatial Context in Recognition. *Perception* **25**, 343–352 (1996).

39. Harel, A. & Bentin, S. Stimulus type, level of categorization, and spatial-frequencies utilization: implications for perceptual categorization hierarchies. *J. Exp. Psychol. Hum. Percept. Perform.* **35**, 1264 (2009).

40. Schyns, P. G. & Oliva, A. Dr. Angry and Mr. Smile: When categorization flexibly modifies the perception of faces in rapid visual presentations. *Cognition* **69**, 243–265 (1999).

41. Beck, D. M. & Kastner, S. Stimulus context modulates competition in human extrastriate cortex. *Nat. Neurosci.* **8**, 1110–1116 (2005).

42. Chen, A. J.-W. *et al.* Goal-directed attention alters the tuning of object-based representations in extrastriate cortex. *Front. Hum. Neurosci.* **6**, 187 (2012).

43. Gazzaley, A., Cooney, J. W., McEvoy, K., Knight, R. T. & D'esposito, M. Top-down enhancement and suppression of the magnitude and speed of neural activity. *J. Cogn. Neurosci.* **17**, 507–517 (2005).

44. O'Craven, K. M., Downing, P. E. & Kanwisher, N. fMRI evidence for objects as the units of attentional selection. *Nature* **401**, 584–587 (1999).

45. Murray, S. O. & Wojciulik, E. Attention increases neural selectivity in the human lateral occipital complex. *Nat. Neurosci.* **7**, 70–74 (2004).

46. Desimone, R. & Duncan, J. Neural mechanisms of selective visual attention. *Annu. Rev. Neurosci.* **18**, 193–222 (1995).

47. Miller, E. K., Freedman, D. J. & Wallis, J. D. The prefrontal cortex: categories, concepts and cognition. *Philos. Trans. R. Soc. Lond. B. Biol. Sci.* **357**, 1123–1136 (2002).
48. Çukur, T., Nishimoto, S., Huth, A. G. & Gallant, J. L. Attention during natural vision warps semantic representation across the human brain. *Nat. Neurosci.* **16**, 763–770 (2013).
49. Toneva, M., Stretcu, O., Póczos, B., Wehbe, L. & Mitchell, T. M. Modeling task effects on meaning representation in the brain via zero-shot meg prediction. *Adv. Neural Inf. Process. Syst.* **33**, 5284–5295 (2020).
50. La Tour, T. D., Eickenberg, M., Nunez-Elizalde, A. O. & Gallant, J. L. Feature-space selection with banded ridge regression. *NeuroImage* **264**, 119728 (2022).
51. Smith, F. W. & Goodale, M. A. Decoding visual object categories in early somatosensory cortex. *Cereb. Cortex* **25**, 1020–1031 (2015).
52. Caspers, S., Zilles, K., Laird, A. R. & Eickhoff, S. B. ALE meta-analysis of action observation and imitation in the human brain. *Neuroimage* **50**, 1148–1167 (2010).
53. Caglar, L. R., Walbrin, J., Akwayena, E., Almeida, J. & Mahon, B. Z. Object-directed action representations are componentially built in parietal cortex. *Proc. Natl. Acad. Sci.* **122**, e2421032122 (2025).
54. Cant, J. S. & Goodale, M. A. Attention to form or surface properties modulates different regions of human occipitotemporal cortex. *Cereb. Cortex* **17**, 713–731 (2007).
55. Peelen, M. V. & Downing, P. E. Selectivity for the Human Body in the Fusiform Gyrus. *J. Neurophysiol.* **93**, 603–608 (2005).
56. Schwarzlose, R. F., Baker, C. I. & Kanwisher, N. Separate face and body selectivity on the fusiform gyrus. *J. Neurosci.* **25**, 11055–11059 (2005).
57. Weiner, K. S. & Grill-Spector, K. Not one extrastriate body area: using anatomical landmarks, hMT+, and visual field maps to parcellate limb-selective activations in human lateral occipitotemporal cortex. *Neuroimage* **56**, 2183–2199 (2011).

58. Vingerhoets, G. Contribution of the posterior parietal cortex in reaching, grasping, and using objects and tools. *Front. Psychol.* **5**, 151 (2014).

59. Garcea, F. E. *et al.* Domain-specific diaschisis: lesions to parietal action areas modulate neural responses to tools in the ventral stream. *Cereb. Cortex* **29**, 3168–3181 (2019).

60. Chao, L. L. & Martin, A. Representation of manipulable man-made objects in the dorsal stream. *Neuroimage* **12**, 478–484 (2000).

61. Mahon, B. Z. *et al.* Action-related properties shape object representations in the ventral stream. *Neuron* **55**, 507–520 (2007).

62. He, K., Zhang, X., Ren, S. & Sun, J. Deep Residual Learning for Image Recognition. in 770–778 (2016).

63. Xiao, S. *et al.* C-Pack: Packed Resources For General Chinese Embeddings. in *Proceedings of the 47th International ACM SIGIR Conference on Research and Development in Information Retrieval* 641–649 (Association for Computing Machinery, New York, NY, USA, 2024). doi:10.1145/3626772.3657878.

64. Dirani, J. & Pylkkänen, L. The time course of cross-modal representations of conceptual categories. *Neuroimage* **277**, 120254 (2023).

65. Gibson, J. J. The theory of affordances. *Hilldale USA* **1**, 67–82 (1977).

66. Yee, E. & Thompson-Schill, S. L. Putting concepts into context. *Psychon. Bull. Rev.* **23**, 1015–1027 (2016).

67. Mahon, B. Z. What is embodied about cognition? *Lang. Cogn. Neurosci.* **30**, 420–429 (2015).

68. Nastase, S. A. *et al.* Attention selectively reshapes the geometry of distributed semantic representation. *Cereb. Cortex* **27**, 4277–4291 (2017).

69. Kanwisher, N. & Wojciulik, E. Visual attention: Insights from brain imaging. *Nat. Rev. Neurosci.* **1**, 91–100 (2000).

70. David, S. V., Hayden, B. Y., Mazer, J. A. & Gallant, J. L. Attention to Stimulus Features Shifts Spectral Tuning of V4 Neurons during Natural Vision. *Neuron* **59**, 509–521 (2008).

71. Mazer, J. A. & Gallant, J. L. Goal-Related Activity in V4 during Free Viewing Visual Search: Evidence for a Ventral Stream Visual Salience Map. *Neuron* **40**, 1241–1250 (2003).
72. Stuss, D. T. & Benson, D. F. The Frontal Lobes and Control of Cognition and Memory. in *The Frontal Lobes Revisited* (Psychology Press, 1987).
73. Badre, D. & Nee, D. E. Frontal cortex and the hierarchical control of behavior. *Trends Cogn. Sci.* **22**, 170–188 (2018).
74. Damen, D. *et al.* Rescaling Egocentric Vision: Collection, Pipeline and Challenges for EPIC-KITCHENS-100. *Int. J. Comput. Vis.* **130**, 33–55 (2022).
75. Damen, D. *et al.* Scaling Egocentric Vision: The EPIC-KITCHENS Dataset. Preprint at https://doi.org/10.48550/arXiv.1804.02748 (2018).
76. CVAT. ai Corporation. Computer Vision Annotation Tool (CVAT). Zenodo https://doi.org/10.5281/ZENODO.12771595 (2024).
77. Zhang, F. *et al.* MediaPipe Hands: On-device Real-time Hand Tracking. Preprint at https://doi.org/10.48550/arXiv.2006.10214 (2020).
78. Nunez-Elizalde, A. O., Deniz, F., Dupré la Tour, T., Visconti di Oleggio Castello, M. & Gallant, J. L. Pymoten: scientific Python package for computing motion energy features from video. *Zenodo* (2021).
79. Cox, R. W. AFNI: software for analysis and visualization of functional magnetic resonance neuroimages. *Comput. Biomed. Res.* **29**, 162–173 (1996).
80. Cox, R. W. & Hyde, J. S. Software tools for analysis and visualization of fMRI data. *NMR Biomed.* **10**, 171–178 (1997).
81. Fischl, B. FreeSurfer. *Neuroimage* **62**, 774–781 (2012).
82. Gao, J. S., Huth, A. G., Lescroart, M. D. & Gallant, J. L. Pycortex: an interactive surface visualizer for fMRI. *Front. Neuroinformatics* **9**, 23 (2015).
83. Huth, A. G., De Heer, W. A., Griffiths, T. L., Theunissen, F. E. & Gallant, J. L. Natural speech reveals the semantic maps that tile human cerebral cortex. *Nature* **532**, 453–458 (2016).

84. Lancaster, J. L. *et al.* Automated labeling of the human brain: A preliminary report on the development and evaluation of a forward-transform method. *Hum. Brain Mapp.* **5**, 238–242 (1997).

85. Lancaster, J. L. *et al.* Automated Talairach Atlas labels for functional brain mapping. *Hum. Brain Mapp.* **10**, 120–131 (2000).

86. Dupré la Tour, T., Visconti di Oleggio Castello, M. & Gallant, J. L. The Voxelwise Encoding Model framework: A tutorial introduction to fitting encoding models to fMRI data. *Imaging Neurosci.* **3**, imag_a_00575 (2025).

87. Kriegeskorte, N., Mur, M. & Bandettini, P. A. Representational similarity analysis - connecting the branches of systems neuroscience. *Front. Syst. Neurosci.* **2**, (2008).

88. van den Bosch, J. J. *et al.* A python toolbox for representational similarity analysis. *bioRxiv* 2025–05 (2025).

## AUTHOR'S NOTE

This work was supported by National Institutes of Health grants R01EY028535 and R01NS089069 to BZM. JD was supported by the Carnegie Mellon University Neuroscience Institute Distinguished Postdoctoral Fellowship. BZM is an inventor of US patents US12437878B2 and US12178591B2, and is a cofounder and Chief Science Officer of MindTrace Technologies Inc which licenses US12437878B2 from Carnegie Mellon University. The authors are grateful to Leyla Caglar for sharing materials for the 20 reference hand synergies.

## DATA AVAILABILITY

Data associated with this paper will be made publicly available upon acceptance for publication.

## SUPPLEMENTAL MATERIALS

| Objects | Actions |
|---|---|
| bacon | close |
| bottle | cut |
| lid | grasp |
| carton/canister | lift |
| colander | mix/stir |
| counter | open |
| mug | peel |
| cupboard | place/put |
| drawer | pour |
| fork | rinse |
| fridge | scoop |
| jar | scrub/wipe |
| knife | turnOff/On |
| knob | |
| microwave | |
| onion | |
| oven mitts | |
| stove | |
| oven-stove | |
| wrapper | |
| pan | |
| pasta | |
| pizza | |
| pizza box | |
| plate | |
| pot | |
| rack | |
| scissors | |
| sink | |
| skin/peel | |
| spatula | |
| sponge/rag | |
| spoon | |
| trashcan | |
| sauce | |

**Table S1.** All unique object and action labels.

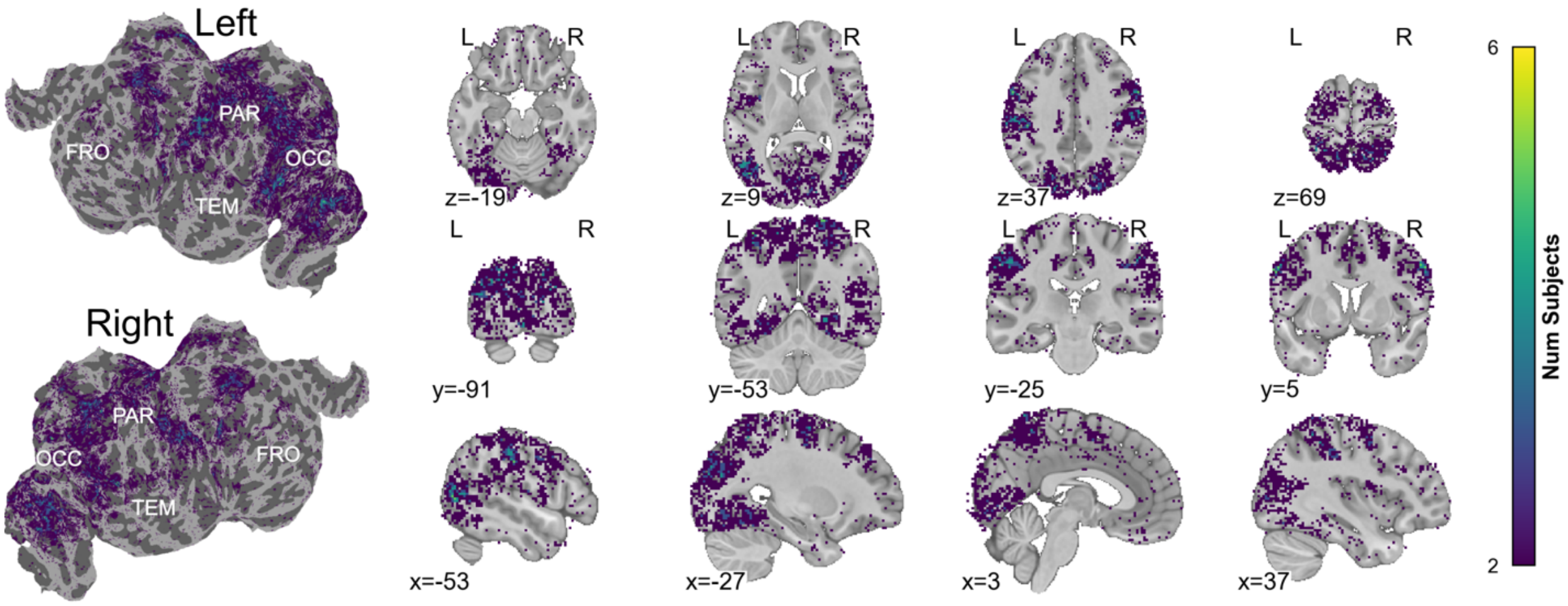

**Figure S2. Subject-level coverage of the voxel-wise encoding model.** Number of subjects for whom voxel-wise model performance was significant ($R^2 > 0$) shown on inflated cortical surfaces (left and right hemispheres; OCC: occipital, PAR: parietal, TEM: temporal, FRO: frontal) and on axial, coronal, and sagittal slices of the MNI template. Significance was established independently within each subject using permutation testing (1,000 iterations, FDR-corrected $q < 0.05$). Voxels with significant model performance in fewer than 2 subjects are kept transparent. This map reflects subject-specific significance independently of the group-level FDR threshold used in **Figure 2**, and shows strong overlap with those results.

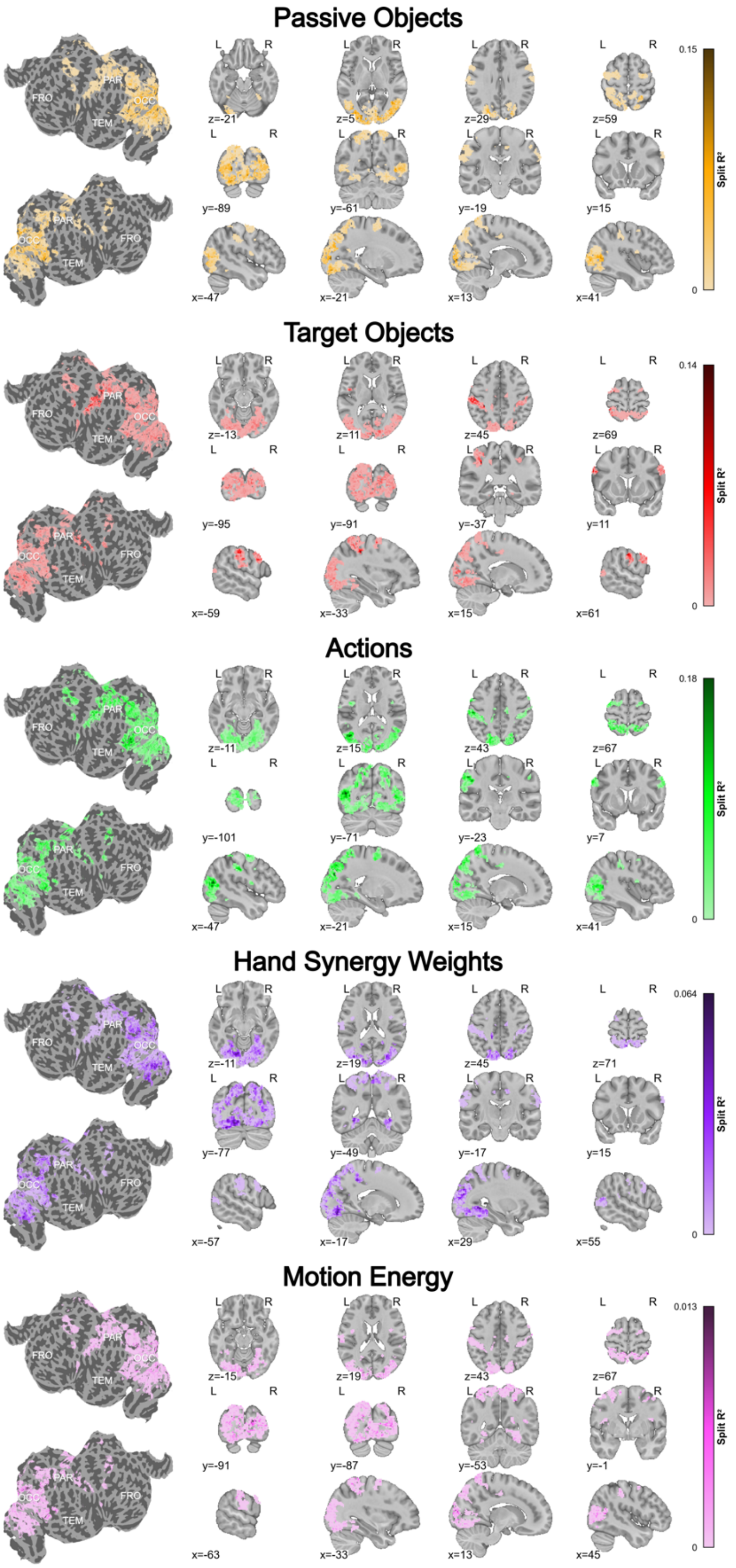

**Figure S3. Whole brain maps of the product-measure split scores ($R^2$) for each of the five feature space.** Voxels not surviving group-level FDR correction ($q < 0.05$, minimum cluster extent 10 voxels) are kept transparent.

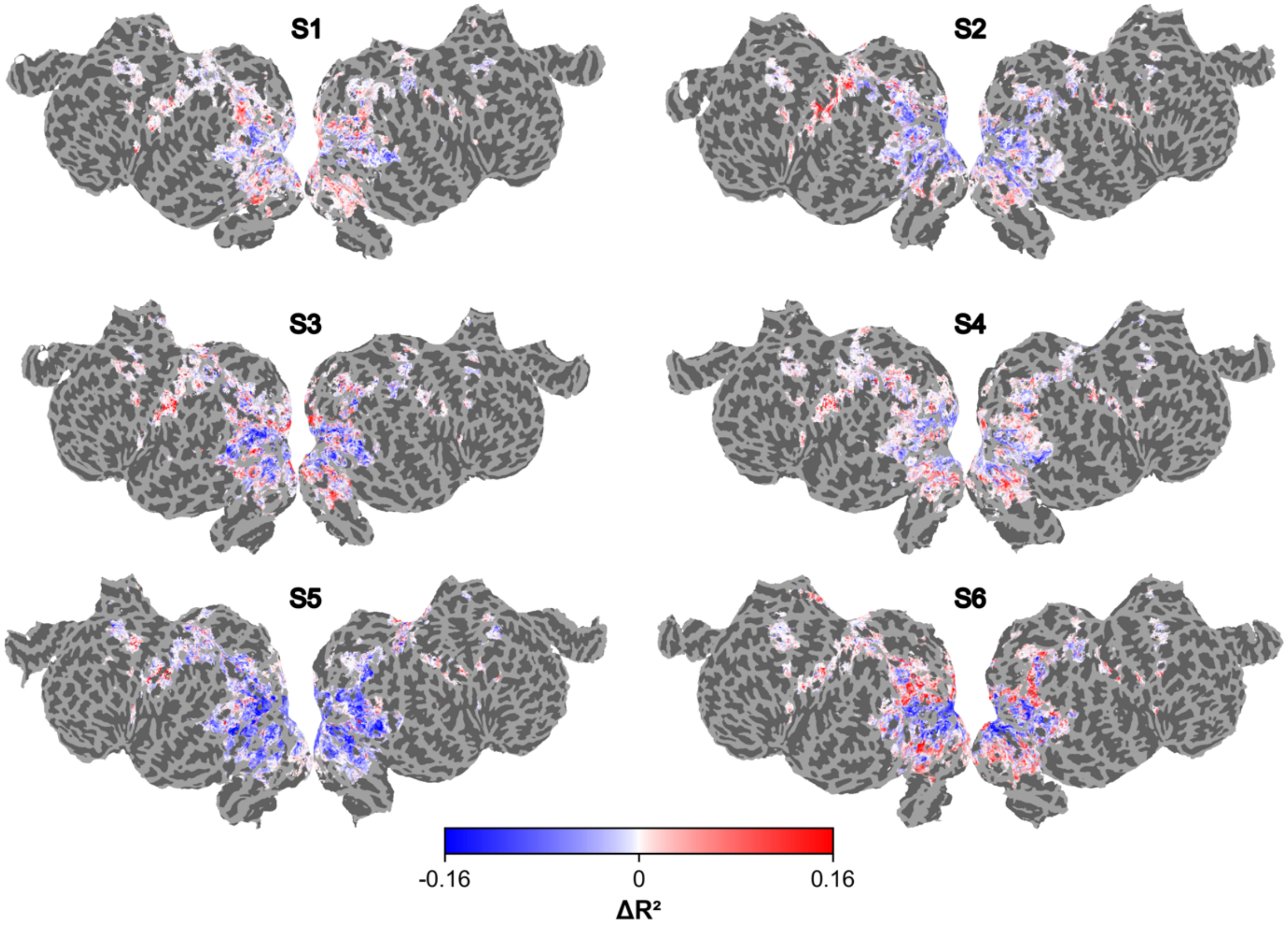

**Figure S4. Subject-level contrast maps displayed within the group-level $R^2$ mask and plotted on the flattened surface, confirming the directional dissociation is present across all six participants**

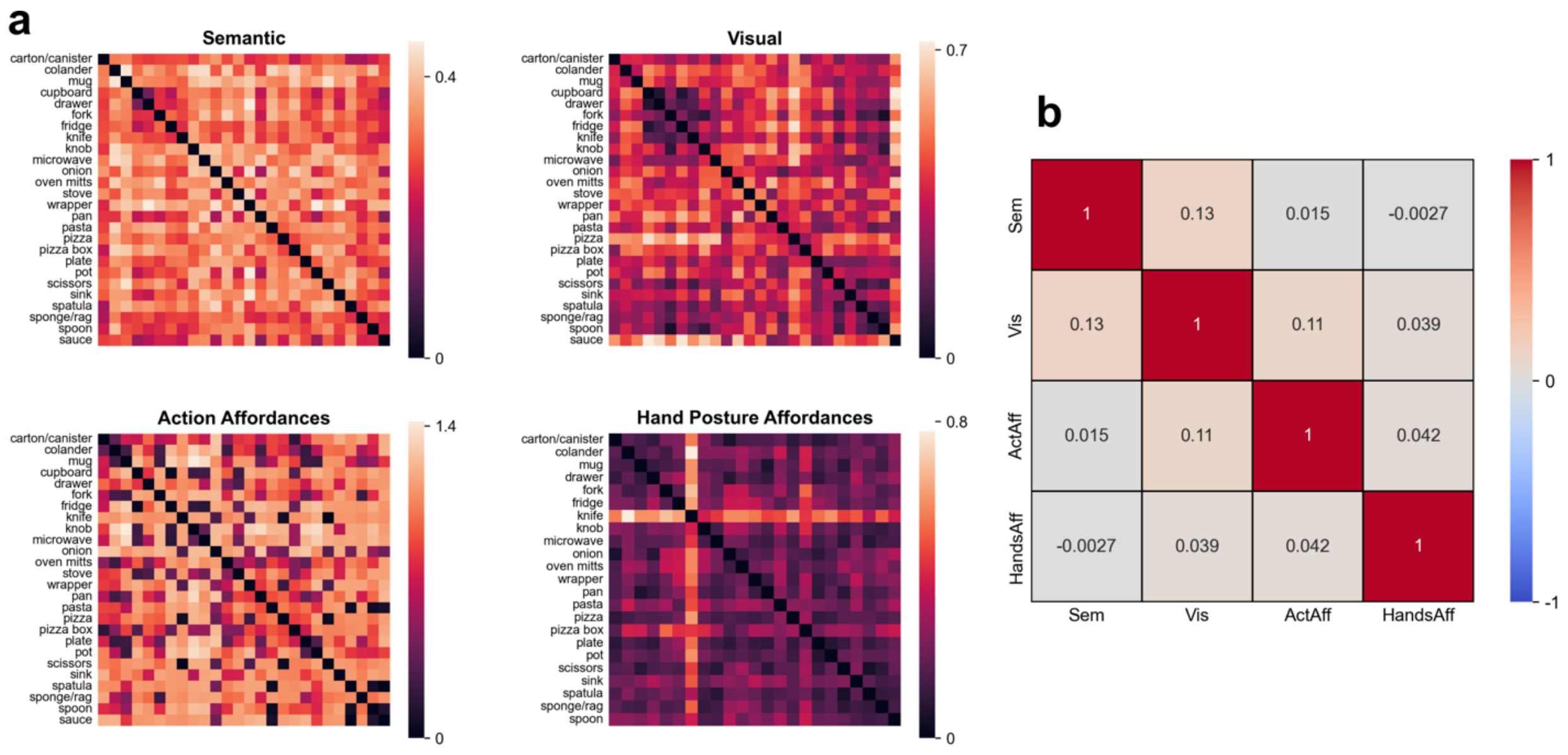

**Figure S5. Hypothesis representational dissimilarity matrices and pairwise correlations**. **(a)** Representational dissimilarity matrices (RDMs) for each of the four hypothesis models across the 26 overlapping objects, computed as pairwise correlation distances (1 − Pearson r) between model embeddings. **(b)** Pairwise Spearman correlations between hypothesis RDMs. Maximum pairwise correlation was r = 0.13 (semantic and visual), confirming that the four hypothesis RDMs capture largely independent dimensions of object representations.

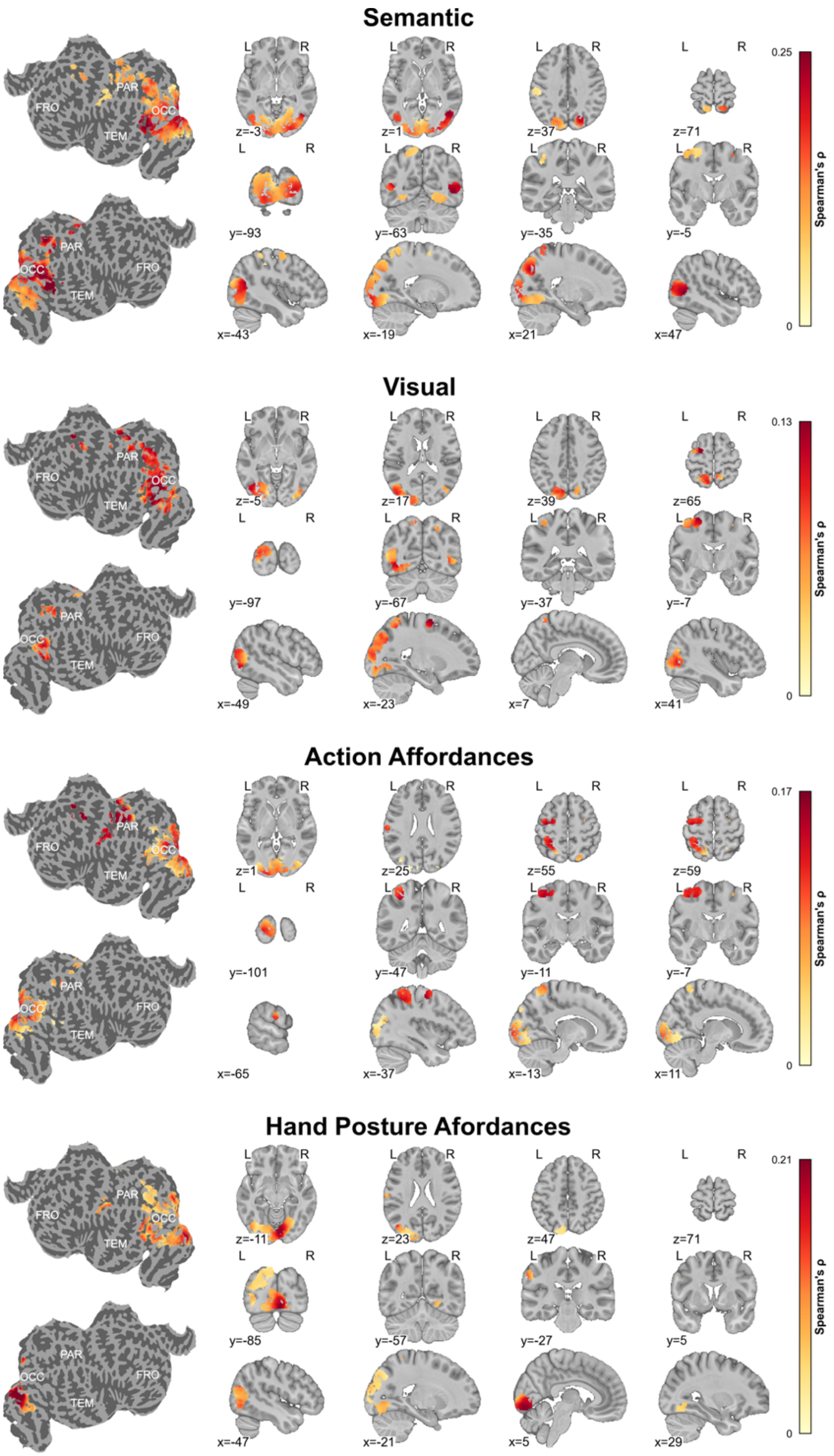

**Figure S6. Searchlight RSA correlation maps for passive objects.** Whole-brain correlation maps showing local alignment between passive object representational geometry and each hypothesis. Voxels not surviving group-level FDR correction ($q < 0.05$, minimum cluster extent 10 voxels) are kept transparent.

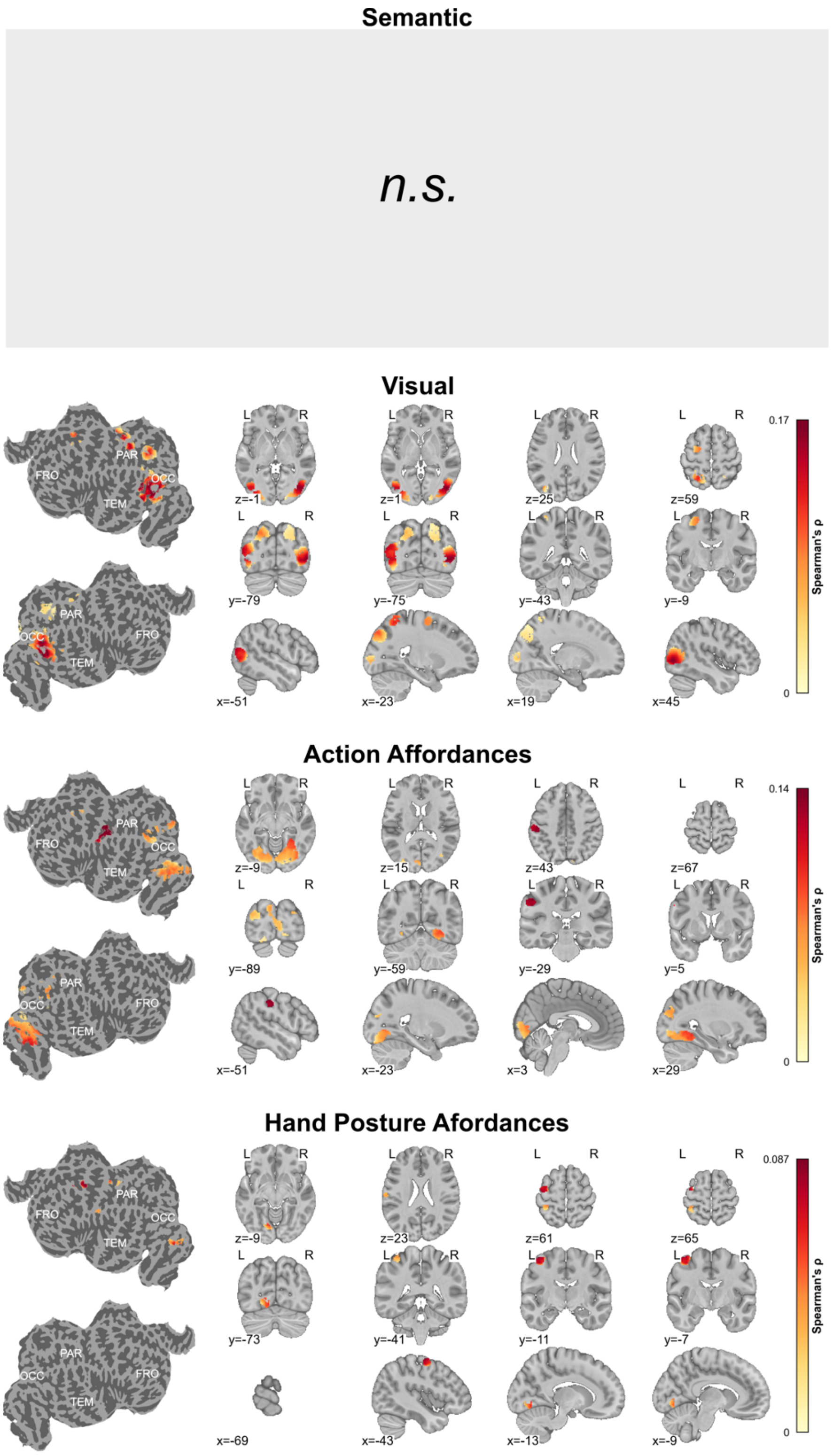

**Figure S7. Searchlight RSA correlation maps for target objects.** Whole-brain correlation maps showing local alignment between target object representational geometry and each hypothesis. Voxels not surviving group-level FDR correction ($q < 0.05$, minimum cluster extent 10 voxels) are kept transparent.

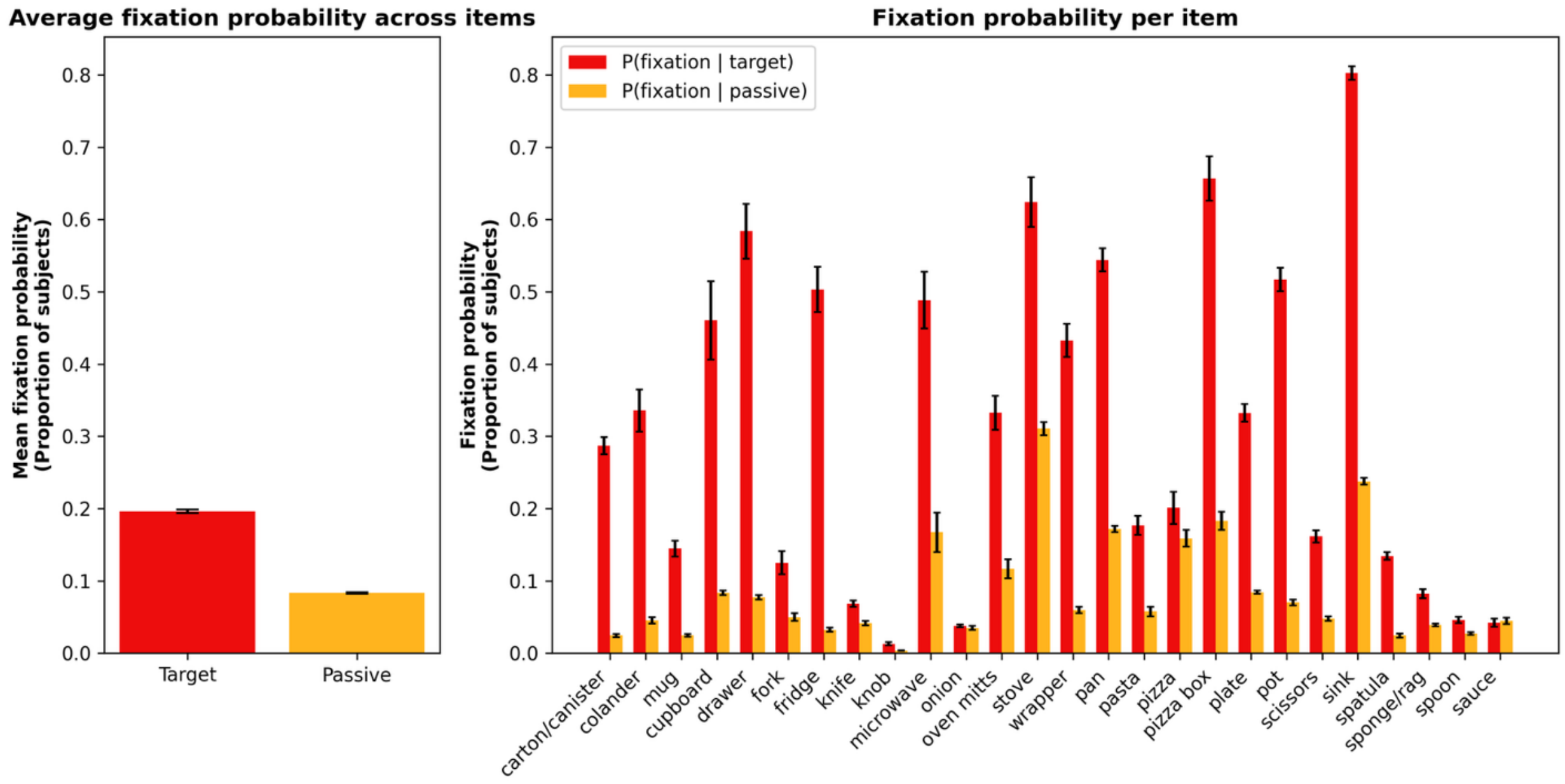

**Figure S8. Fixation probability for target and passive objects.** Results from an independent eye-tracking experiment (N = 34) in which participants viewed the same movie stimuli outside the scanner. Fixation probability was computed per frame as the proportion of participants whose gaze fell within the annotated bounding box of an object. The left panel shows mean fixation probability averaged across all 26 objects, separately for target and passive contexts. The right panel shows mean fixation probability for each object across all frames in which they appear, separately for target (red) and passive (orange) contexts.

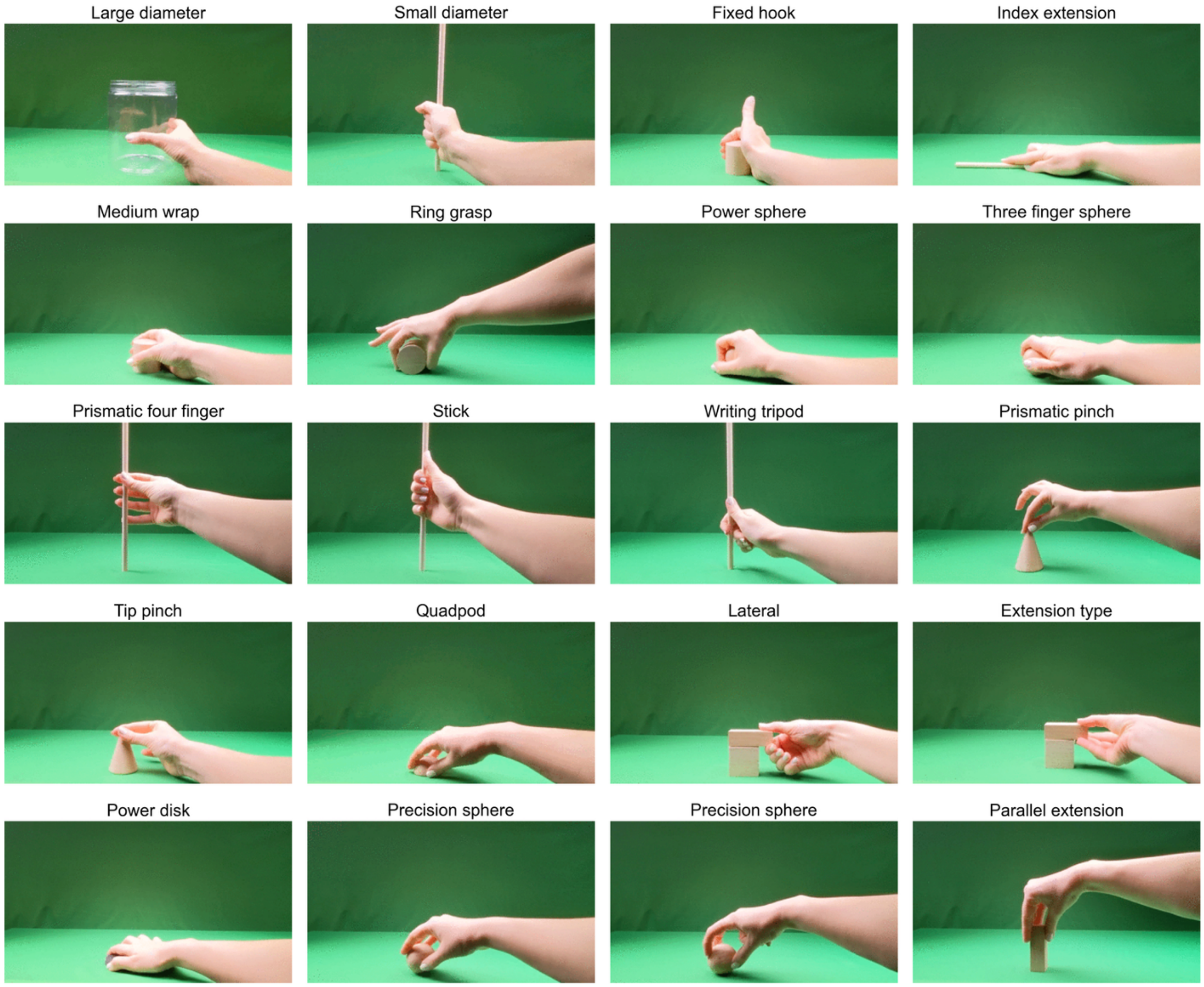

**Figure S9.** The 20 reference hand synergies from Caglar et al. (2025) used to decompose hand configurations in the movie stimuli. Each panel shows the canonical photograph of a hand posture corresponding to one synergy. Movie hand skeletons were expressed as weighted linear combinations of these 20 synergies, yielding the hand synergy weights used as a feature space in the voxel-wise encoding model.